\documentclass[12pt,preprint]{emulateapj}
\usepackage{epsfig}
\usepackage{graphicx}
\def\gta{\mathrel{\hbox{\rlap{\hbox{\lower4pt\hbox{$\sim$}}}\hbox{$>$}}}}

\shorttitle{SFR in z=1 Clusters}
\shortauthors{Krick et al.}

\begin{document}
\newcommand\msun{\hbox{M$_{\odot}$}}
\newcommand\lsun{\hbox{L$_{\odot}$}}
\newcommand\magarc{mag arcsec$^{-2}$}
\newcommand\h{$h_{70}^{-1}$}

\bibliographystyle{apj}
\title{\bf Galaxy Clusters in the IRAC Dark Field II: \\ Mid-IR Sources}

\author{J.E.~Krick \altaffilmark{1}, J.A.~Surace \altaffilmark{1}, D.~Thompson \altaffilmark{2}, M.L.N.~Ashby \altaffilmark{3}, J.L.~Hora \altaffilmark{3}, V.~Gorjian \altaffilmark{4}, and L.~Yan \altaffilmark{1}}

\altaffiltext{1}{Spitzer Science Center, MS 220--6,
California Institute of Technology, Jet Propulsion Laboratory,
Pasadena, CA 91125, USA}
\altaffiltext {2}{Large Binocular Telescope Observatory, University of Arizona, Tucson, AZ 85721}
\altaffiltext {3}{Harvard-Smithsonian Center for Astrophysics, 60 Garden Street, Cambridge MA 02138}
\altaffiltext {4}{Jet Propulsion Laboratory, California Institute of Technology, Pasadena, CA, 91109}
\email{jkrick@caltech.edu}

\begin{abstract} 

  We present infrared luminosities, star formation rates (SFR),
  colors, morphologies, locations, and AGN properties of 24\micron\
  -detected sources in photometrically detected high-redshift clusters
  in order to understand the impact of environment on star formation
  and AGN evolution in cluster galaxies.  We use three
  newly-identified $z=1$ clusters selected from the IRAC dark field;
  the deepest ever mid-IR survey with accompanying, 14 band
  multiwavelength data including deep {\it HST} imaging and deep
  wide-area {\it Spitzer} MIPS 24 micron imaging. We find 90 cluster
  members with MIPS detections within two virial radii of the cluster
  centers, of which 17 appear to have spectral energy distributions
  (SED) dominated by active galactic nuclei (AGN) and the rest
  dominated by star formation.  We find that 43\% of the star forming
  sample have infrared luminosities $L_{IR} > 10^{11} \lsun$ (luminous
  infrared galaxies; LIRGs).  The majority of sources (81\%) are
  spirals or irregulars.  A large fraction (at least 25\%) show
  obvious signs of interactions.  The MIPS -detected member galaxies
  have varied spatial distributions as compared to the MIPS-undetected
  members with one of the three clusters showing SF galaxies being
  preferentially located on the cluster outskirts, while the other 2
  clusters show no such trend.  Both the AGN fraction and the summed
  SFR of cluster galaxies increases from redshift zero to one, at a
  rate that is a few times faster in clusters than over the same
  redshift range in the field.  Cluster environment does have an
  effect on the evolution of both AGN fraction and SFR from redshift
  one to the present, but does not effect the infrared luminosities or
  morphologies of the MIPS sample.  Star formation happens in the same
  way regardless of environment making MIPS sources look the same in
  the cluster and field, however the cluster environment does
  encourage a more rapid evolution with time as compared to the field.

\end{abstract}


\section{Introduction} 
\label{intro}
Galaxy groups and clusters represent the dense environments required for
hierarchical galaxy formation.  Cluster galaxies potentially follow a
different evolutionary path from isolated field galaxies because of a cluster's
large gravitational potential and hot, X-ray emitting gas.  As galaxy
clusters form and grow throughout time by infall of galaxies and
groups of galaxies, the simple picture is one of member galaxies that
are transformed from blue, late-types with signs of star formation to
red, early-types with no star formation.  This conversion most likely
happens through a combination of processes including mergers, star
formation bursts, ram pressure stripping, and harassment
\citep{vandokkum2005, gunn1972, moore1996}.

This work comes at a key time in the study of star forming galaxies
and AGN in high-redshift clusters.  Only recently have we been able to
study star formation in clusters at $z=1$.  There are only a few
well-studied clusters at $z=1$, although the number is growing rapidly
and will continue to do so with upcoming Sunyaev Zeldovich and large
sky surveys \citep{staniszewski2008}.  Additionally, traditional
measures of star formation are difficult to obtain at high redshifts.
H$\alpha$ shifts out of the optical band by $z\sim 0.5$.  Both OII and
H$\alpha$ narrow band surveys with specially designed filters
\citep{poggianti2008,finn2008} are possible, but optical emission line
spectroscopy at high redshift is telescope time intensive, and narrow
band surveys only work for the designed redshift.  Both of these
measures are also affected by dust obscuration. However, with {\it
  Spitzer} MIPS we are able to measure obscured star formation at
large redshifts with relative ease.

That we see star formation in galaxy clusters at all is worth
investigation.  O \& B stars live for less than 10 million years, so a
single, triggered episode of star formation is likely to last for less
than few tens of million years.  If the infall time of a galaxy into
the center of a cluster is roughly 1Gyr (assuming 1 Mpc radius and
1000km/s velocities) and all galaxies somehow have their star
formation suppressed upon entering the cluster environment, we would
expect to see {\it no} star formation in the centers of clusters,
unless it is triggered, in situ, by mergers.  We would therefore
expect to see no blue, isolated galaxies with heightened star
formation in the central regions of clusters.  Furthermore, if star
formation is actually first triggered and then suppressed upon cluster
entry, as it has been suggested processes like ram pressure stripping
could do \citep{bekki2003, kronberger2008}, then we should see star
formation in isolated spirals on the outskirts of the clusters.  Based
on this timescales argument we should potentially see star formation
on the edges of clusters, but not in the centers, unless it is merger
driven. Star formation triggered by galaxy interactions and mergers is
not dependent on cluster environment, instead on the relative
velocities of the galaxies.  As such this form of star formation can
occur anywhere in the cluster environment or the field, and is more
likely to happen in lower mass clusters or groups due to the lower
relative velocities.

There is intriguing evidence that star formation rates in clusters
increase with redshift out to at least $z=0.83$ \citep{bai2007}.  We
investigate if this evolution follows that in the field, implying that
cluster environment does not impact star formation.  We examine this
claim by increasing the number of clusters studied at high redshift
and extending the redshift range out to redshift one.  There are only
two clusters with published MIPS 24\micron\ star formation rates at
redshifts above 0.8 , both at $z=0.83$ \citep[MS1054-03, RX
J0152][]{bai2007,marcillac2007,saintonge2008}.  \citet{koyama2008} use
the Infrared Camera on Akari \citep{onaka2007, murakami2007} at
15\micron s to study a redshift 0.81 cluster.  Although this is a
mid-IR measurement of SFR, they use a different rest-frame wavelength
to convert to $L_{IR}$ which carries a different set of assumptions.
Our survey is unique in that we double the number of published high
redshift clusters with 24\micron\ star formation rates by adding a
large scale structure at $z=1$ containing three clusters/groups with
larger number statistics and deeper $L_{IR}$ measurements over a large
area.

In addition to star formation, we examine for the first time
MIPS-detected AGN in cluster environments as a different line of
evidence of galaxy activity.  The same processes which affect star
formation in galaxies will also effect the AGN on roughly the same
timescales \citep{hopkins2008}.  AGN and star formation are linked not
only because they both require a cold gas reservoir to ignite, but
also due to both star formation an AGN feedback mechanisms which have
the ability to destroy or remove the cold gas and halt either the star
formation, the AGN activity, or both \citep{croton2006}.  AGN can put
a halt to star formation by blowing out or heating the gas, and
similarly star formation can use up gas thereby removing the source
for a central engine.  We expect the AGN fraction at high redshift to
be higher than at low redshift in clusters based on evidence both in
clusters and the field \citep{osmer2004, eastman2007, kocevski2008,
  galametz2009}.  We examine if the AGN fraction in clusters tracks
the redshift evolution of that in the field or is enhanced by the
cluster environment.

This paper is structured in the following manner. In
\S\ref{observations} \& \S\ref{photz} we discuss the data and derived
photometric redshift determination.  Details of the sample selection
are presented in \S\ref{sample}.  In \S\ref{results} we present the
AGN fraction, infrared luminosities, star formation rates, colors,
morphologies, and radial distributions of both the star forming and
AGN samples. The paper is summarized and conclusions are drawn in
\S\ref{conclusion}.  Throughout this paper we use $H_0=70$km/s/Mpc,
$\Omega_M$ = 0.3, $\Omega_\Lambda$ = 0.7.  With this cosmology, the
luminosity distance at z=1 is 6607 Mpc, but the angular diameter
distance is a factor of $(1+z)^2$ less, or 1652 Mpc.  All photometry
is quoted in the AB magnitude system.


\section{Observations \& Data Reduction}
\label{observations}

\subsection{The IRAC Dark Field}

The survey region is the IRAC Dark Field, centered at approximately
17h40m +69d.  The field is located a few degrees from the north
ecliptic pole (NEP) in a region which is darker than the actual pole
and is in the {\it Spitzer} continuous viewing zone so that it can be
observed any time IRAC is powered on for observing.  These observing
periods are called instrument ``campaigns'', and occur roughly once
every three to four weeks and last for about a week. Sets of long
exposure frames are taken on the Dark Field at least twice during each
campaign totaling roughly four hours of integration time per campaign,
and these data are used to derive dark current/bias frames for each
channel.  The dark frames are used by the pipeline in a manner similar
to ``median sky'' calibrations as taken in ground-based near-infrared
observing to produce the Basic Calibrated Data (BCD) for all science
observations. Each set of dark calibration observations collects
roughly two hours of integration time at the longest exposure times in
each channel.

The resulting observations are unique in several ways.  The Dark Field
lies near the lowest possible region of zodiacal background, the
primary contributor to the infrared background at these wavelengths,
and as such is in the region where the greatest sensitivity can be
achieved in the least amount of time.  The area was also chosen
specifically to be free of bright stars and very extended galaxies,
which allows clean imaging to very great depth. The observations are
done at many position angles (which are a function of time of
observation) leading to a more uniform final point spread function
(PSF).  Finally, because the calibration data are taken directly after
anneals, they are more free of artifacts than ordinary guest observer
(GO) observations.  Over the course of the mission, the observations
have filled in a region 20\arcmin \ in diameter with a total of $\sim
350$ hours devoted to the project; $\sim 70$ hours per pixel in the
center of each band as of the writing of this paper.  This will create
the deepest mid-IR survey, exceeding the depth of the deepest planned
regular {\it Spitzer} surveys over several times their area.
Furthermore, this is the only field for which a 5+year baseline of
mid-IR periodic observations is expected.

The IRAC data is complemented by imaging data in 14 other bands with
facilities including Palomar, MMT, {\it HST}, Akari, {\it Spitzer}
MIPS, and Chandra ACIS-I as well as Palomar optical spectroscopy.
Although the entire dark field is $> 20\arcmin$ in diameter, because
of spacecraft dynamics the central $\sim 15\arcmin$ is significantly
deeper and freer of artifacts.  Therefore, it is this area which we
have matched with the additional observations.  The entire dataset
will be presented in detail in a future paper (Krick et al, in prep).
For completeness we briefly discuss here the {\it Spitzer} IRAC, {\it
  Spitzer} MIPS, {\it HST} ACS, and Palomar optical spectroscopy as
they are the most critical to this work.  All space-based datasets are
publicly available through their respective archives.

\subsection{Spitzer IRAC}
\label{irac}

This work is based on a preliminary combination of 75 hours of IRAC
imaging, which is $\approx$30\% of the expected depth not including
the warm mission.  The Basic Calibrated Data (BCD) product produced by the
{\it Spitzer} Science Center was further reduced using a modified version of
the pipeline developed for the SWIRE survey \citep{surace2005}.  This
pipeline primarily corrects image artifacts and forces the images onto
a constant background (necessitated by the continuously changing
zodiacal background as seen from {\it Spitzer}). The data were coadded onto
a regularized $0\farcs6$ \ grid using the {\sc mopex} software developed by
the {\it Spitzer} Science Center.

Experiments with DAOPHOT demonstrate that nearly all extragalactic
sources are marginally resolved by IRAC, particularly at the shorter
wavelengths, and hence point source fitting is inappropriate.
Instead, photometry is done using the high spatial resolution ACS data
as priors for determining the appropriate aperture shape for
extracting the Spitzer data.  We do this by first running source
detection and photometric extraction on the coadded IRAC images using
a matched filter algorithm with image backgrounds determined using the
mesh background estimator in SExtractor (Bertin et al. 1995) .  This
catalog is merged with the HST ACS catalog.  For every object in that
catalog ,if the object is detected in ACS then we use the ACS shape
parameters to determine the elliptical aperture size for the IRAC
images.  ACS shape parameters are determined by SExtractor on
isophotal object profiles after deblending, such that each ACS pixel
can only be assigned to one object (or the background).  For objects
which are not detected in ACS, but which are detected in IRAC, we
simply use the original IRAC SExtractor photometry.  Because of the
larger IRAC beam, we impose a minimum semi-major axis radius of
2\arcsec.  In all cases aperture corrections are computed individually
from PSF's provided by the SSC based on the aperture sizes and shapes
used for photometry.

Final aperture photometry was performed using custom extraction
software written in IDL and based on the APER and MASK$\_$ELLIPSE
routines with the shape information from SExtractor, from either ACS
or IRAC as described above, using local backgrounds.  Because we use
local backgrounds, the measured fluxes of objects near the confusion
limit should have a larger scatter than those non-confused objects,
but will on average be the correct flux.  This will not effect the
photometric redshifts, as it will likely shift all IRAC points up or
down, but not relative to each other.  

Determining the detection limits of the IRAC data is complicated by
varying exposure times across the field, source confusion, and our use
of ACS locations as priors for photometry.  Because of these three
complexities, there is no one single value for the detection limit of
the survey, however this work is limited by the MIPS detection limits
and not IRAC or ACS.  We measure nominal $95\%$ completeness limits in
the IRAC passbands from a number count diagram at 3.6, 4.5, 5.8, and
8.0\,\micron\ to be 0.2, 0.17, 0.11, 0.11\,$\mu$Jy respectively.

\subsection{Spitzer MIPS24}
\label{MIPS24}
The {\it Spitzer} MIPS 24\micron\ data were taken in large-field
photometry mode with a 30-second exposure time. A $3\times3$ MIPS
field of view grid was mapped and repeated five times, with multiple
dithers and chops totaling 224 sq. arcminutes in the center of the
IRAC image. There were a total of 1080 separate exposures with a final
total depth of 60 minutes per pointing on the sky.  The MIPS data were
processed by the {\it Spitzer} Science Center into individual image
BCDs. However, substantial ``jailbar'' artifacts, as well as a
significant gradient, were visible. All of the frames were forced to a
common background by applying an additive constant to the entire
frame. A ``delta-dark'' was then generated from the median of all
frames; the great degree of dithering in the data allows this process
to reject all actual celestial objects in the frames from the median
stack. That stack was then adjusted to a median overall zero value,
and then subtracted from all the data. It currently is not known
whether the gradient effect is additive or multiplicative, although
our experience with other Si:As arrays of this kind strongly suggests
(from a physical basis) that it is additive. However, we reduced the
data both ways, and found no difference at any detectable level. The
data were then coadded using the {\sc mopex} software package onto
exactly the same projection system as used for IRAC, albeit with
$1\farcs2$ pixels.

IRAF {\sc daofind} was used for object detection.  We supply the code
with the PSF FWHM and background sigma values taken by examining the
image. {\sc Daofind} then counts the flux within an annulus of
diameter FWHM and flags any set of pixels as a detection where that
flux is above a threshold of five sigma.  To deal with confused
sources, we perform object detection iteratively.  After the first run
through {\sc daofind}, all objects are subtracted from the image using
a PSF determined from the detected objects. {\sc Daofind} is then
re-run on the residual image.  To ensure that the iterative detection
is not dominated by noise, we manually check all detections within the
cluster area by eye (see \S \ref{sample}).  With the exception of a
handful of galaxies, all MIPS detections appear as point sources.
Photometry on all detected sources is done with the IRAF task {\sc
  allstar} which fits PSF's to groups of objects simultaneously.  An
aperture correction of 1.4 is applied for flux beyond the 6.5 pixel
radius at which the PSF star was normalized.  This correction factor
is calculated from a curve of growth based on the composite PSF star.
Using this method the $3\sigma$ detection limit is 17.3$\mu$Jy.  These
noise properties are comparable to the GOODS slightly longer exposure
(77 minute) dataset that has a $3\sigma$ limit of 12$\mu$Jy.

\subsection{HST ACS}
\label{acs}
The {\it HST} observations consist of 50 orbits with the ACS
comprising 25 separate pointings, all with the F814W filter (observed
I-band).  Within each pointing eight dithered images were taken for
cosmic ray rejection and to cover the gap between the two ACS CCDs.
The ACS pipeline {\sc calacs} was used for basic reduction of the
images.  Special attention was paid to bias subtraction, image
registration, and mosaicing.  Pipeline bias subtraction was
insufficient because it does not measure the bias level individually
from each of the four amplifiers used by ACS.  We make this correction
ourselves by subtracting the mean value of the best fit Gaussian to
the background distribution in each quadrant.  Due to distortions in
the images, registration and mosaicing was performed with a
combination of IRAF's {\sc tweakshifts, multidrizzle,} and {\sc SWarp}
  v.2.16.0 from Terapix.  The actual task of mosaicing the final
image was complicated by the large image sizes.  The single combined
mosaic image is 1.7GB and reading in all 200 images (160Mb each) for
combination is impossible for most software packages.

The final combined ACS image is $\sim 15\arcmin$ diameter coincident
with the deepest part of the IRAC Dark Field and is made with the
native $0.05\arcsec$ per pixel resolution.  Photometry was performed
in a standard manner with {\sc  SExtractor}. The $3\sigma$ detection limit
for point sources is $F814W = 28.6(AB)$.

\subsection{Palomar Optical Spectroscopy}
\label{spectroscopy}

The Palomar data consists of a total of four nights at the Hale 200''
telescope with the COSMIC spectrograph.  COSMIC, at prime focus, has a
13.6\arcmin\ field of view, and 0.4\arcsec\ pixels.  Observations were
made on a total of four photometric nights in June of 2007 \& 2008
with the 300 l/mm grating with a dispersion of 2 \AA\ per pixel.  We
chose a slit-width of 1.5\arcsec\ to match our 1 - 1.5\arcsec\ seeing.
The optical band covered by this instrument includes such spectral
features as CaH\&K, [OII], [OIII], H$\alpha$, H$\beta$, H$\delta$, G
band, and the 4000 \AA break.  During both runs we were able to
observe a total of 11 slitmasks of $\sim 25$ galaxies each with
exposure times of on average 80 minutes divided into multiple
exposures.  One Hg-Ar lamp and one flat was taken through each mask at
the beginning of the night for calibration.  Galaxies were chosen to
be brighter than r=21(AB) with priority given to those with MIPS 24 or
70\micron\ detections to boost the chance of seeing an emission line
and thereby getting a secure redshift.

Reduction was done with IRAF mainly through the {\sc Bogus2006}
\footnote{https://zwolfkinder.jpl.nasa.gov\slash ~stern\slash
  homepage\slash bogus.html} scripts.  Prior to running bogus, images
were overscan and bias subtracted.  {\sc Bogus} itself does a 2D reduction
including flat-fielding, cosmic ray removal, sky subtraction, fringe
suppression and combination of frames.  The same reduction is performed
on both science images and arcs.  The standard IRAF tasks of {\sc apall,
identify}, and {\sc dispcor} were used to wavelength correct, trace, and
extract the spectra with a secondary background subtraction for minor
level changes.  One dimensional spectra were extracted for a total of 200 galaxies
with measurable continuum.

No single cluster galaxy was bright enough to have a spectrum observed
at Palomar.  Instead these spectra are used to calibrate our
photometric redshifts.


\section{Photometric redshifts}
\label{photz}

The combined IRAC and ACS catalog contains over $50,000$ objects which
makes acquisition of spectroscopic redshifts impractical.  Even
confirmation spectroscopy of red galaxies at $z=1$ in our three
candidate clusters will require many nights on 8-10m class telescopes
and is therefore also impractical.  In lieu of spectroscopy we use our
extensive multi-wavelength, broad-band catalog to build spectral
energy distributions (SEDs) using up to 13 bands (u', g', r', i',
F814W, z', J, H, K, 3.6, 4.5, 5.8, 8.0\micron) from which we derive
photometric redshifts .  A full discussion of the accuracy of
photometric redshift determinations is beyond the scope of this paper
\citep[but see for example
][]{mobasher2004,brodwin2006,bolzonella2000}.

These SEDs are fit with template spectra derived from galaxies in the
{\it Spitzer} wide area infrared survey survey
\citep[SWIRE;][]{Polletta2007}.  These templates have been used
successfully by a number of surveys at a range of redshifts for all
galaxy types \citep{adami2008,negrello2008,salvato2009,ilbert2009}.
Since the SWIRE templates are based on {\it Spitzer} observations we
find them the best choice to use as models for this dataset.  We use
15 templates including ellipticals, spirals, star forming galaxies,
and AGN.  Photometric redshifts are calculated using {\sc Hyperz}; a
chi-squared minimization fitting program including a correction for
interstellar reddening \citep{bolzonella2000, calzetti2000}.

Errors in photometric redshifts are determined by comparing the
photometric redshifts with spectroscopic redshifts.  Spectroscopic
redshifts were determined using both IRAF tasks {\sc emsao} and {\sc
  xcsao}.  Specifically {\sc emsao} searches the spectrum for both
absorption and emission lines which it correlates with a given line
list.  {\sc xcsao} cross-correlates the spectrum with known galaxy
templates which allows us to use features like the 4000 \AA\ break and
the rest of the spectral shape to identify redshifts.  Both techniques
were used together to arrive at the best fit redshift for each
galaxy. We used 17 spectral templates of galaxies and AGN from the
compilation of the {\it HST} Calibration Database System
\citep{francis1991, kinney1996, calzetti1994}.  We applied a very
strict requirement that all emission and absorption features in the 1D
spectra were confirmed by eye on the 2D spectra and that multiple
lines be identified in all cases to avoid incorrect redshift
determination due to cosmic rays or noise features from sky line
subtraction.

\begin{figure}
\epsscale{0.5}
\plotone{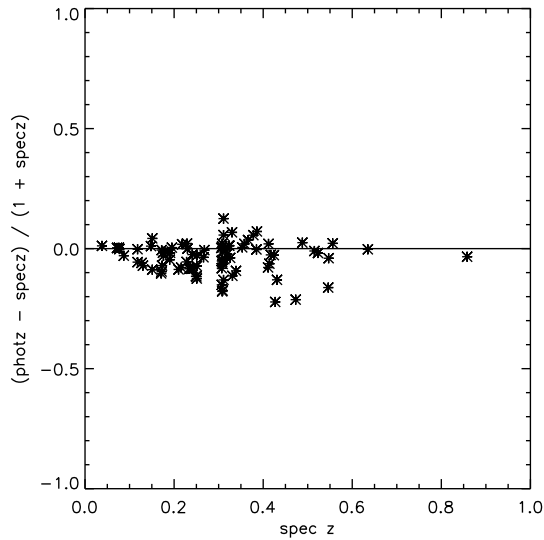}
\caption[specvsphotz]{Comparison of spectroscopically and
  photometrically determined redshifts.  The scatter implies an error on the photometric
  redshifts of $0.064(1+z)$.}
\label{fig:specvsphotz}
\epsscale{1}
\end{figure}

We were able to successfully determine redshifts for 87 galaxies.
This represents a conservative sample of 'good' redshift
determinations defined to have either high signal-to-noise emission
lines or multiple absorption lines and good cross correlations.  We
then compare the spectroscopic to photometric redshifts to quantify
the error on the latter (Figure \ref{fig:specvsphotz}). There are
cases where {\sc Hyperz} has failed to fit the correct redshift which
is obvious when looking at the SED fit.  Those galaxies, as
characterized by a $\chi^{2}$ value greater than 50, are not included
in this comparison or the cluster sample below.  The error on the
photometric redshifts is $0.064(1+z)$.  Note that this error is quoted
as a function of redshift and so takes into account the increasing
scatter with z.  This accuracy is similar to other IRAC based
multi-wavelength studies \citep{brodwin2006, rowanrobinson2008}.  We
are confident that our quoted accuracy will hold in extrapolating our
photmetric redshifts out to z=1 because at that redshift the Balmer
break is shifted into our HST ACS F814W and MMT $z'$ which are our
most sensitive bands.  Secondly the peak of the stellar distribution
is shifted into the IRAC bands where we have excellent coverage. It
should be noted that while this level of accuracy is standard, it
still implies a large volume at z=1 and therefore our sample selection
below likely includes foreground and background interlopers.  We have
no leverage to remove these without exhaustive spectroscopic data.

\section{Sample Selection}
\label{sample}

A detailed description of the cluster properties, masses, color
magnitude diagrams, and redshift distributions is given in paper one
\citep{krick2008}. Table \ref{tab:clusterchar} is reproduced here from
that paper to summarize their properties. The first cut we make on the
sample is that the objects need to have detections in at least six
bands to ensure that they are real detections and not noise
fluctuations. Because we use ACS locations to measure IRAC fluxes,
there are cases where ACS noise (diffraction spikes, etc.) will get
picked up as an object with five flux measurements. On the other hand
there are real cluster galaxies which are only detected in ACS + IRAC
bands because ACS is the deepest band blue-ward of IRAC and the SED's
are falling sharply into the blue.

We choose twice the virial radius as the interesting physical radius
that includes the dense core of the cluster but also the infall region
out to roughly the turnaround radius where we might expect to find
different populations of galaxies. Cluster centers are determined from
the spatial distribution of the member galaxies in the F814W filter.
We determine the virial radius from our X-ray detections (see Paper 1,
Figure 3 for the Chandra image). The diffuse Chandra detections give
us $r_{500}$; the radius at which the cluster has 500 times the
critical density of the Universe. From there we derive $r_{vir}$
assuming that $r_{500} = 0.6*r_{vir}$ \citep{johnston2007}. This
relation between $r_{500}$ and $r_{vir}$ comes from the average of
130,000 groups and clusters from SDSS. For our relatively low mass
clusters $r_{vir}$ is 0.7 ,0.58, and 0.58 Mpc, which corresponds to
$87\farcs3 ,72\farcs8$, and $72\farcs8$ respectively. Cluster two \&
three are too close to discuss separately as their virial radii are
overlapping. We therefore consider them as one structure. The
selection area will be the addition of the two circular regions. For
cluster one we only consider half of the possible total area because
the other half is not completely covered by our ACS imaging. While the
ACS data is missing, we do have IRAC and MIPS data for this region
which indicates that the cluster is symmetric and therefore we
are not missing an obvious population by cutting the cluster in half.

Cluster members are chosen by their photometric redshifts.  The
cluster redshift distributions are centered at z=1.0.  Our photometric
errors at this redshift are 0.13, so we take as members all galaxies
within $0.87 < z_{phot} < 1.13$ with {\sc Hyperz} chi-squared values
less than 50.  This high value cutoff of chi-squared is to keep out
the catastrophic failures of {\sc Hyperz}.  We do not use the red
sequence to determine membership because we expect some of the member
galaxies to be blue, particularly those with MIPS detections, and we
don't want to bias this work against those galaxies.

Overall there are 443 member galaxies with detections in at least six
bands and positions within two virial radii of any of the cluster
centers, 90 of those have 24\micron\ detections with $f_{24} >
17.3\mu$Jy.  Because the PSF of MIPS is larger than the IRAC PSF, we
checked by eye all MIPS matches for all objects within the area of the
clusters to ensure that the correct matches with the closest centers
were chosen.  In the case of ambiguity, where multiple galaxies could
have matched the MIPS source, those sources were not included in the
analysis (approximately 10 sources).  We also checked by eye those
MIPS sources that were not determined to be members to make sure that
a mis-match did not occur that would have kept those objects out of
the member list.  This fraction of members with MIPS detections of
20\% is in the right ballpark when compared to those in the literature
given the varying methods of determining membership, varying depths,
and different cluster masses.  \citet{bai2007} find that $13\pm3\%$ of
cluster members are actively forming stars with $f_{24} > 50\mu$Jy.

Because we have a relatively large area at redshift one in the IRAC
dark field, we are also able to make a redshift one 'field' sample of
those galaxies with the exact same criteria as above (secure
detections, z=1, and $f_{24} > 17.3$) except that they are required to
be {\it more} distant than two viral radii of the cluster centers.

\section{Results \& Discussion}
\label{results}

\subsection{Dominant SED Shape}
\label{selection}

Because infrared flux can be generated either by dust re-radiating
young star light or accretion onto a black hole, we attempt to divide
the sample into sources where the MIPS flux is likely to be dominated
by star formation and those where an AGN likely dominates.  There is no
perfect way to determine this division and it is very likely that
sources have signatures of both processes (see \S \ref{intro}).  The
best discriminator for the available data are the differing spectral
shapes of the UV to mid-IR range for AGN and galaxies.  AGN have
red continuua in this range owing to their rising power law shape as
opposed to the falling blackbody in the same wavelength regime for
galaxies.  We choose to use the SED shapes as fitted by {\sc Hyperz}
to determine if the source spectrum is best fit by a star forming
galaxy or an AGN.

\begin{figure}
\epsscale{0.5}
\plotone{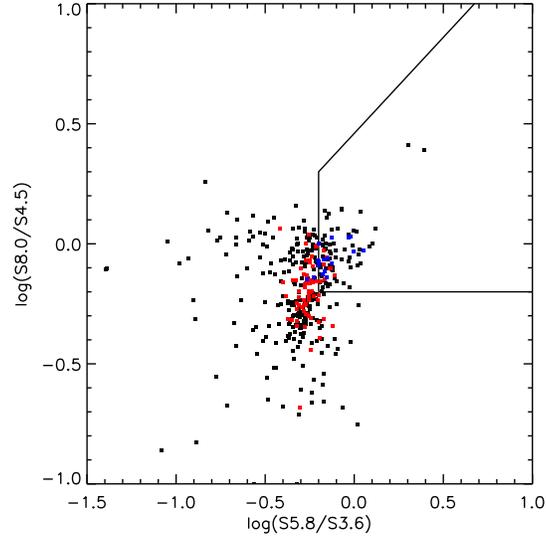}
\caption[iraccolor]{IRAC color color diagram after \citet{lacy2004}.
  All cluster member galaxies are shown in black.  Those with
  24\micron\ detections are color coded red for star forming and blue
  for AGN based on {\sc Hyperz} fits of their SEDs.  Lines show the
  expected location of AGN based on having red colors in both
  axes.  The member galaxies on the color color diagram are where we
  expect redshift one galaxies to be \citep{sajina2005} based on
  position of PAH features and the stellar peak, which is a nice
  confirmation of our photometric redshifts.}
\label{fig:iraccolor}
\epsscale{1}
\end{figure}

AGN candidates account for 17 of the 90 member galaxies with
24\micron\ detections or 19\% of MIPS sources and 4\% of all members.
These are referred to in the rest of the paper as the AGN sample.  The
remaining 76 galaxies have SEDs which are dominated by star formation
and are referred to here as the star forming member sample.  Figure
\ref{fig:iraccolor} shows the IRAC color-color diagram for all member
galaxies as a complementary method of separating AGN from star forming
galaxies \citep{lacy2004, stern2005}.  Those galaxies with MIPS
24\micron\ detections are denoted with red (star forming galaxy) or
blue (AGN) colors based on their {\sc Hyperz} fits.  It is
unsurprising to find that the sources tagged as AGN by their spectral
fits also fit into the AGN wedge with 88\% completeness but with
significant contamination; 40\%.  The contamination is likely from
intermediate redshift, PAH dominated galaxies and is similar in
quantity to simulations by \citet{sajina2005}.

\subsection{AGN fraction}

We compare here the evolution of AGN dominated MIPS sources in
clusters with that in the field.  These are the first AGN fractions of
MIPS-detected sources in clusters at high redshift.  MIPS is sensitive
to the compton thick AGN not detectable at other wavelengths.  The
literature does hold published X-ray-based AGN fractions in clusters.
The only other IR work on this topic was published very recently
by \citet{galametz2009} based on observed frame IRAC colors and not mid-IR
luminosities.  Both the X-ray and near-IR studies find tantalizing
evidence for an increasing AGN fraction with increasing redshift
\citep{martini2007,eastman2007,kocevski2008, cappelluti2005}.  In a
compilation, \citet{eastman2007} look at the redshift evolution of the
AGN fraction where AGN were selected from a sample of cluster members
with $M_{R} < -20$.  X-ray point sources with luminosities above
$1\times10^{43} erg/s$ were counted in comparison to the member
galaxies.  They find a trend of AGN fraction increasing from 0.07\% to
2\%, an increase of a factor of $\sim20$, over the redshift range $0.2
< z < 0.6$.  \citet{kocevski2008} look at a similar sample of X--ray
sources in a supercluster at $z=0.9$ and confirm the trend of higher
AGN fraction at these higher redshifts.  Although we have {\it
  Chandra} data which detect diffuse emission from two of the
clusters, we are not able to do a point source analysis in the
clusters to any meaningful depths.

In order to make a similar AGN fraction measurement to those described
above we attempt to make similar flux cuts on our sample to find those
galaxies that could potentially host AGN and the subset of those that
we measure to be AGN dominated.  We take the potential hosts to be all
galaxies at the cluster redshift with $M_{R} < -20$ corresponding to
$m_r < 24.8$, including K and evolution corrections for early-type
galaxies.  Although these sources potentially have non-early-type
SED's, we chose those K and evolutionary corrections to be consistent
with what was done in \citet{eastman2007}.  Instead of a limit on
X-ray luminosity, we use a correlation between $L_x$ and
$L_{5.8\micron}$ \citep{fiore2008} to determine which of our member
AGN have $L_{5.8} > 2.5\times 10^{43}$ ergs/s.  With these
requirements we find there are 97 member galaxies within one virial
radii of the cluster which have $m_r < 24.8$.  Four of those galaxies
have SED shapes of AGN and $L_{5.8} > 2.5\times 10^{43}$ ergs/s or
$4\%$ of the possible hosts.  When compared to the \citet{eastman2007}
fraction of 0.07\% at z=0.2, our data shows an increase f a factor of
60 of the AGN fraction in clusters from redshift 1.0 to 0.2.

We caution that this fraction depends relatively heavily on the
magnitude limit of the sample and the $L_x$-$L_{5.8}$ correlation.  If
we change the magnitude cut to include fainter (brighter) galaxies
down to $m_r < 25.8$ ($m_r < 23.8$) then we find a ratio of 2.5\%
(6.5\%), both of which still represent an increase over lower redshift
clusters but show a large range.  If we use the \citet{lutz2004}
relation for the $L_x$ and $L_{5.8}$ relation where the $L_x > 1\times
10^{43} $ergs/s limit corresponds to $L_{5.8} > 3.5\times 10^{43}$
ergs/s then we find a fraction of 1\%.  We also caution that
interpretations about the existence of a trend in AGN fraction with
redshift are limited by the small number the comparison samples.  At
least for the radio active galaxies, \citet{lin2007} find that the
radio active fraction depends both on the luminosity limit of the
sample and the mass of the cluster, such that more luminous galaxies
and more massive clusters are likely to have higher fractions of radio
active galaxies.  They posit that this is really only an effect of the
luminosity limit since lower mass clusters are also less likely to
have high luminosity galaxies.

We now compare the evolution of AGN density (number of AGN per
Mpc$^3$) in clusters with that in the field. Using the same cluster
samples, sample selection, and caveats as above, we calculate that the
AGN density in clusters evolves by a factor of $\sim500$ from z=1 to
z=0.2. This number is also uncertain for the same reasons as mentioned
in the previous paragraph and will drop to a factor of $\sim 100$ if
the $L_x - L_{5.8}$ relation of \citep{lutz2004} is used. This
measurement of density uses a volume measurement in our data which is
a cylinder with depth equal to our redshift uncertainty and a radius
of r$_{200}$. We compare this AGN density to a field sample of
\citet{ueda2003} which is a compilation of many surveys with AGN
having L$_{X} > 1\times 10^{43}$erg/s. They find only a factor of
$\sim10$ increase in the field AGN density over the same redshift
range. This implies stronger cluster evolution of the AGN density as
compared to the field, or that cluster environment has influence over
AGN evolution. This same trend is also reported in \citet{eastman2007}
and \citet{galametz2009}. The difference in reported strength is
likely due to using different redshift ranges, AGN detection
techniques, brightness cutoffs, and the other caveats mentioned above.

\subsection{Star Formation Rate }
\label{sfr}

Rest-frame 12\micron\ flux correlates with total infrared luminosity
($L_{IR}$) which can then be converted into star formation rate (SFR).
The correlation between 12\micron\ and $L_{IR}$ is due to the PAH
emission lines.  In the absence of longer wavelength data, which is
not possible to get at high enough resolution and sufficient depth for
these clusters, 12\micron\ is the best wavelength from which to make
the conversion; more secure than both 7 and 15\micron\
\citep{chary2001}.  The correlation between $L_{IR}$ and SFR comes
from the interstellar dust that absorbs the UV-optical light of young
stars and re-radiates that energy in the infrared.  With our
24\micron\ flux and photometric redshifts we estimate the total IR
luminosity using the methods of \citet{chary2001}.  Specifically
templates from both those authors and \citet{dale2002} are redshifted
to our source redshift and then matched to the observed 24\micron\
flux.  The best-fitting template from each model is then used to
derive an average $L_{IR}$.  From there we derive the SFR using the
correlation from \citet{kennicutt1998}.  The described conversion from
$L_{12}$ to SFR is uncertain by factors of a few.  However, we note
that many of the conclusions of this paper rely not on the absolute
SFR, but on the detection of some amount of star formation in cluster
galaxies.

\begin{figure}
\epsscale{0.5}
\plotone{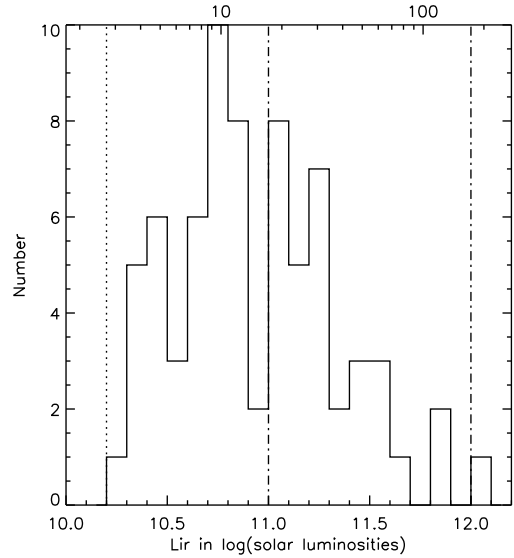}
\caption[lirhist]{Histogram of infrared luminosity of the star forming
  member galaxies.  The top axis shows SFR in \msun/year.  Dot-dashed
  lines show the cutoff for LIRGs and ULIRGs at $1\times 10^{11}$ and
  $1\times 10^{12}$ \lsun respectively. 43\% of the sample are above
  the LIRG cutoff.  The dotted line shows the completeness limit of
  the MIPS data.}
\label{fig:lirhist}
\epsscale{1}
\end{figure}

A histogram of $L_{IR}$ from the star forming member sample are shown
in Figure \ref{fig:lirhist}.  $43\%$ of the sample have infrared
luminosities greater than $1\times10^{11} \lsun$ making them luminous
infrared galaxies (LIRGs).  One galaxy has a flux of $1.01\times
10^{12} \lsun$ qualifying it to be an ultra-luminous
infrared galaxy (ULIRG).  We find a similar ratio of LIRGs to sub-LIRGs as other
clusters at higher redshift.  \citet{marcillac2007} finds 60\% of
their star forming sample (30 galaxies) at $z=0.83$ are LIRGs to a
very similar detection limit.  In a different cluster at $z=0.83$,
\citet{bai2007} finds 41\% of their sample (34 galaxies) are LIRGs.
However, that survey is not as deep which means there will be more
sub-LIRGs which will make this fraction lower.  \citet{geach2006} in
two clusters at $z \sim 0.4\ \&\ 05$ don't go deep enough to get a good
sample of sub-LIRGs.

We compare the luminosity distributions of star-forming cluster
members to field MIPS-detected galaxies at redshift one.  A KS test
between the two distributions shows them to have a $99\%$ probability
of being drawn from the same population.  This would imply that the
cluster environment does not affect the infrared luminosity of the
galaxies within it.  In other words, among star forming galaxies, star
formation does not vary with environment.

In addition to calculating individual star formation rates per galaxy
we compare the total star formation rate per cluster with other
clusters at varying redshifts from the literature.  The interesting
physical quantity to compare is the mass-normalized SFR because SFR
could vary with mass of the cluster \citep[although see][]{goto2005}.
We compare our redshift one clusters with 14 clusters with $0.02 < z <
0.83$ from the literature \citep[][and references therein]{bai2007}.
The literature sample selects only those galaxies with SFR
$>2$\msun/yr within $0.5r_{200}$.  Our SFR cutoff is similar
(~3\msun/yr) and we truncate our sample to match the $0.5r_{200}$
radius.  

\begin{figure}
\epsscale{0.5}
\plotone{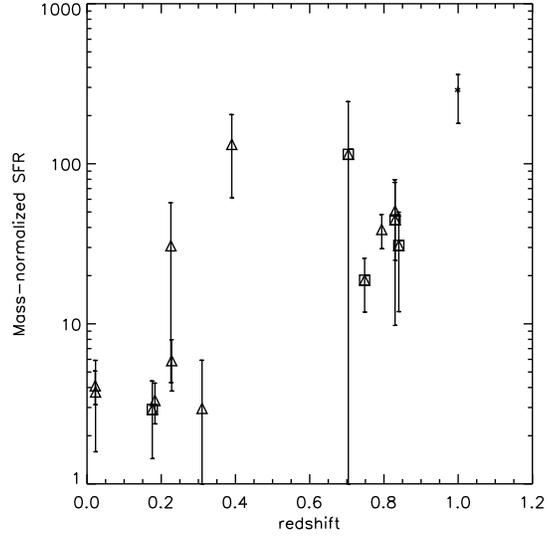}
\caption[massnormsfr]{Mass normalized star formation rate as a
  function of redshift.  The asterisk represents the three redshift
  one clusters from this survey.  The triangles are from the
  literature.  Those literature clusters with masses less than
  $5\times10^{14}$ have their triangles surrounded by squares.  Error
  bars come from a combination of mass and SFR errors.}
\label{fig:massnormsfr}
\epsscale{1}
\end{figure}

In Figure \ref{fig:massnormsfr} the literature clusters are shown with
triangles and the composite of our redshift one clusters with an
asterisk. Error bars in all cases are $1\sigma$ errors taken from the
combination of both mass and SFR errors. Our three clusters are
relatively low mass clusters, and because there is some concern about
a relation between mass-normalized SFR and mass, we also denote the
other lower mass clusters ($M < 5\times10^{14}$\msun) in this figure
with squares. These lower mass clusters in the comparison sample are
still of higher mass than our redshift one clusters. However,
hierarchical formation tells us that redshift one clusters will grow
in mass by the time they reach redshift zero. Comparing clusters of
the same mass across a large redshift range would then also introduce
a bias into the sample. Our three redshift one clusters are suggestive
of continuing the trend of higher redshift clusters having a larger
amount of mass-normalized SFR. This is true both when looking at the
whole sample of lower redshift clusters and also confining the sample
to the five lowest mass, lower redshift clusters. It will be important
to compare our clusters to even lower mass, low-redshift counterparts
when that data becomes available.

SFR can also be computed from different wave-bands.  A detailed
discussion of the varying methods and their relative strengths and
weaknesses is beyond the scope of this paper, but see
\citet{kennicutt1998} and references thereto for such a discussion.
We would like to compare our results on the redshift evolution of the
mass-normalized SFR with other measures from the literature, however
such measures are not published covering the entire redshift range
presented here.  H$\alpha$ and OII derived SFR for clusters at z$ <
0.8$ are presented in \citet{finn2008} and \citet{poggianti2008} and
those are in agreement with the mid-IR determined values
\citep{bai2007}.

The comparison with the field SFR evolution is also interesting.  We
know that the SFR density (SFRD) in the universe peaks around $1< z <
2$ and then declines to today \citep{madau1998, lilly1996}. A recent
compilation of SFRD measurements, \citet{hopkins2006} show a factor of
$5 \sim 10$ drop in the SFRD from redshift 1.0 to 0.1 in the field. We
calculate the SFRD of our clusters at z=1 and compare this to the
SFRDs for the four lowest redshift clusters in the literature sample
with an average redshift of z=0.1.  We do this using the sample
confined to 0.5r$_{200}$ for ease of comparison.  We find a drop in
SFRD from z= 1.0 to z=0.1 of a factor of 40.  This could imply that
while the distribution of IR luminosities of z=1 cluster galaxies are
similar to the field, the suppression of star formation happens more
quickly in clusters than in the field, implying that the cluster
environment is more efficient in the suppression of star formation and
AGN than the field.  Our data suggest this is the case, but a larger,
more uniform sample is required for confirmation.

\subsection{Color}
\label{color}

\begin{figure}
\epsscale{0.4}
\plotone{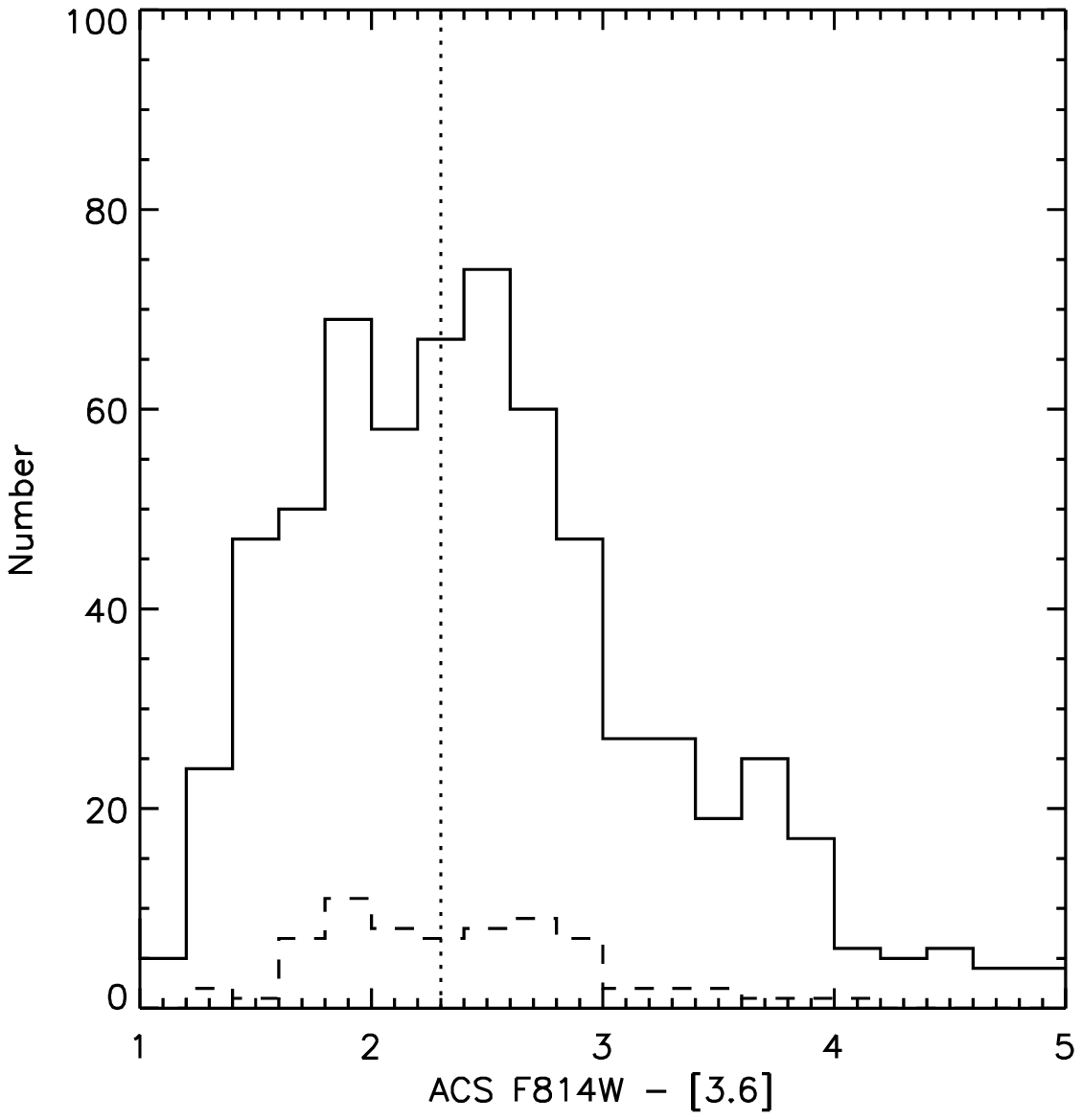}
\plotone{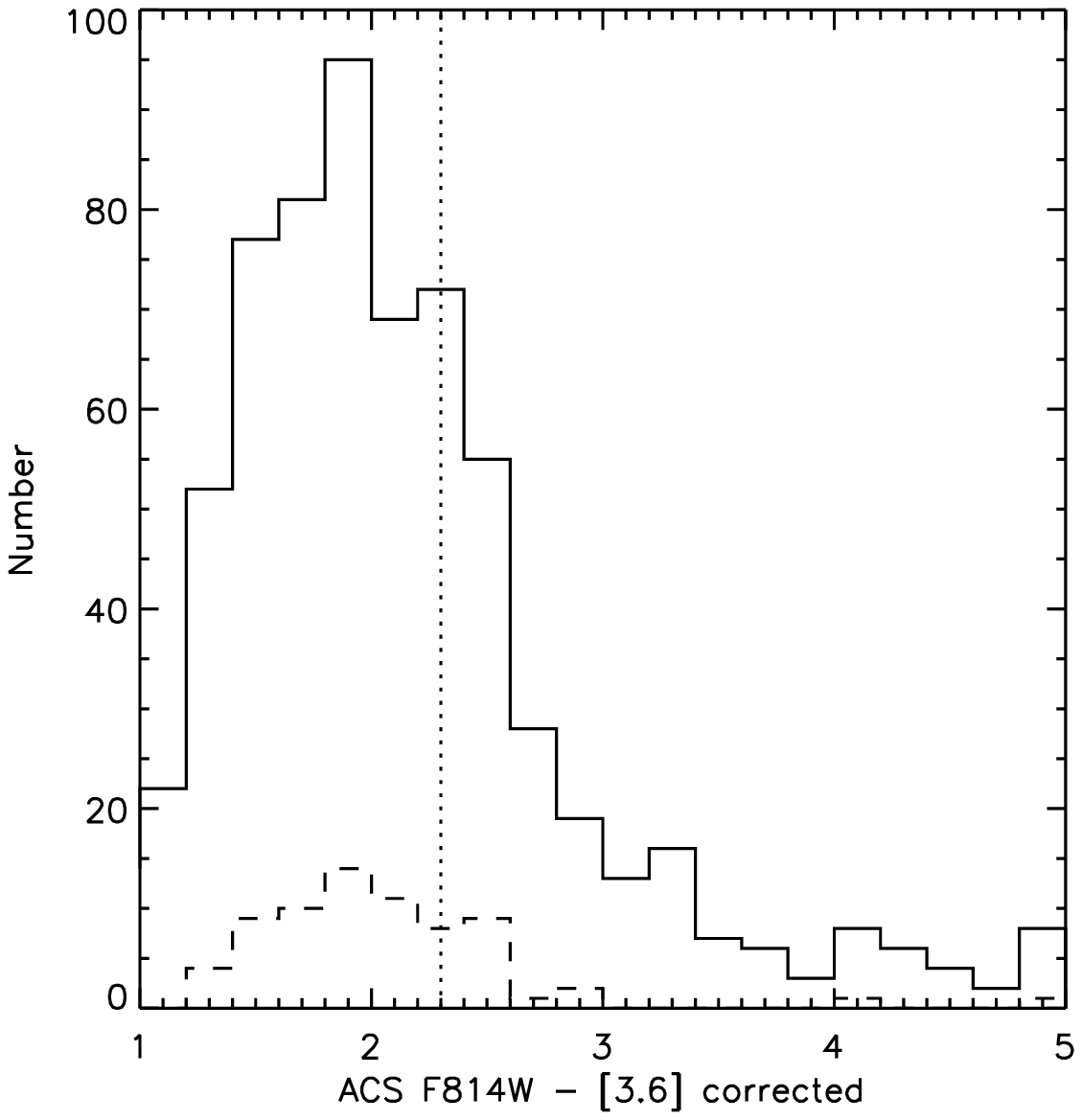}
\caption[colordist]{Histogram of colors of the star forming member
  galaxies (dashed) and all member galaxies (solid).  At redshift one
  this color range is rest-frame B-K.  The vertical dotted line
  shows roughly where the division between red and blue galaxies lies.
  The right figure shows the color distribution of the same samples
  where colors are corrected for extinction based on SED fitting and a
  Calzetti extinction law.}
\label{fig:colordist}
\epsscale{1}
\end{figure}

We explore the colors of the MIPS-detected, star forming sources in
the clusters for the purpose of understanding if the red galaxies in
the clusters are red because they have no star formation, or if they
are red due to dust.  Figure \ref{fig:colordist} shows the distribution
of rest-frame B-K colors of the MIPS-detected, star forming member galaxies
(dashed line) and all cluster members(solid line).
We use the dotted line as the dividing line between the blue cloud
and the red sequence (see Paper 1).

We correct galaxy colors for dust reddening using the extinction as
measured by the {\sc Hyperz} SED fits and the \citet{calzetti2000}
extinction law.  Another possible way to make this correction is with
Balmer line spectroscopy.  However, with a sample of greater than 2000
galaxies at $0.05 < z < 1.5$, \citet{cowie2008} find that SED fitting
is a comparable technique and in fact use the SED fitted extinction
instead of the Balmer ratios even when they do have spectroscopy.  The
corrected colors for our sample are shown on the right side of Figure
\ref{fig:colordist}.

There is a significant amount of extinction at these wavelengths,
particularly at rest-frame B where extinctions range from $A_{B} = 0 -
1.6$, showing that many of these galaxies are dust reddened star
forming sources and in large part not galaxies that are red due to
age.  The corrected histogram shows a very different distribution,
with 57\% of the MIPS sources moving from the red sequence to the blue
cloud.This is consistent with \citet{cowie2008} who find roughly half
of their MIPS-detected red sequence galaxies move off of the red
sequence after correction.  These data tell us that the MIPS sources
do not form a uniquely colored population and are instead very dusty
galaxies.

\subsection{Morphology}
\label{morph}
\begin{figure}
\epsscale{0.5}
\plotone{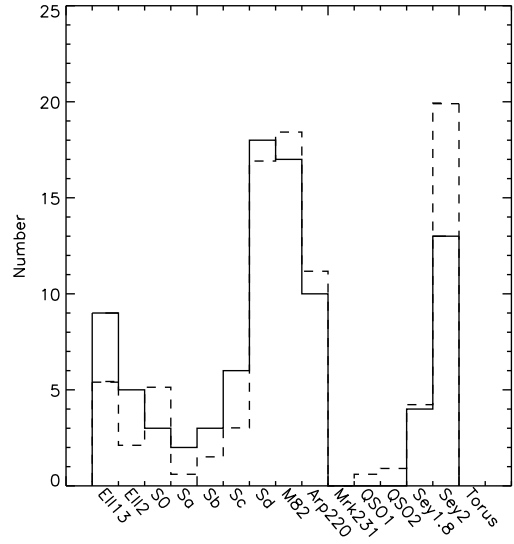}
\caption[hyperzmorph]{Best fit morphologies from {\sc Hyperz} SED fits.  The solid
line shows the member galaxies with 24\micron\ detections and the
dashed histogram is the normalized histogram of all field 24\micron\
detections.}
\label{fig:hyperzmorph}
\epsscale{1}
\end{figure}

We examine morphologies of the MIPS sample both with SED fitting and a
by-eye determination for the purpose of determining which types of
galaxies are mid-IR bright in clusters at $z=1$.  One method of
determining galaxy type is by fitting templates to it's SED.  This
really is a measure of the SED shape used as a proxy for morphology.
The strength of this method is that it allows us to easily compare
field to cluster galaxies using the same objective criteria.  SED shape has
already been determined for all galaxies with {\sc Hyperz} while fitting for
photometric redshifts.  In Figure \ref{fig:hyperzmorph} we show the
histogram of types of galaxies from this analysis arranged from star
forming galaxies to AGN.  The solid line shows the member galaxies
with 24\micron\ detections and the dashed histogram is the normalized
histogram of all 24\micron\ detections across the entire field.  As
expected there are relatively few early-type galaxies, and a
relatively large number of late-types and AGN.  There are very similar
distributions from cluster to field.  This is perhaps hinting that
cluster environment is not effecting the morphologies of the mid-IR
bright galaxies, much like the infrared luminosities of star forming
galaxies being unaffected by environment in \S \ref{sfr}.

\begin{figure}
\epsscale{0.3}
\plotone{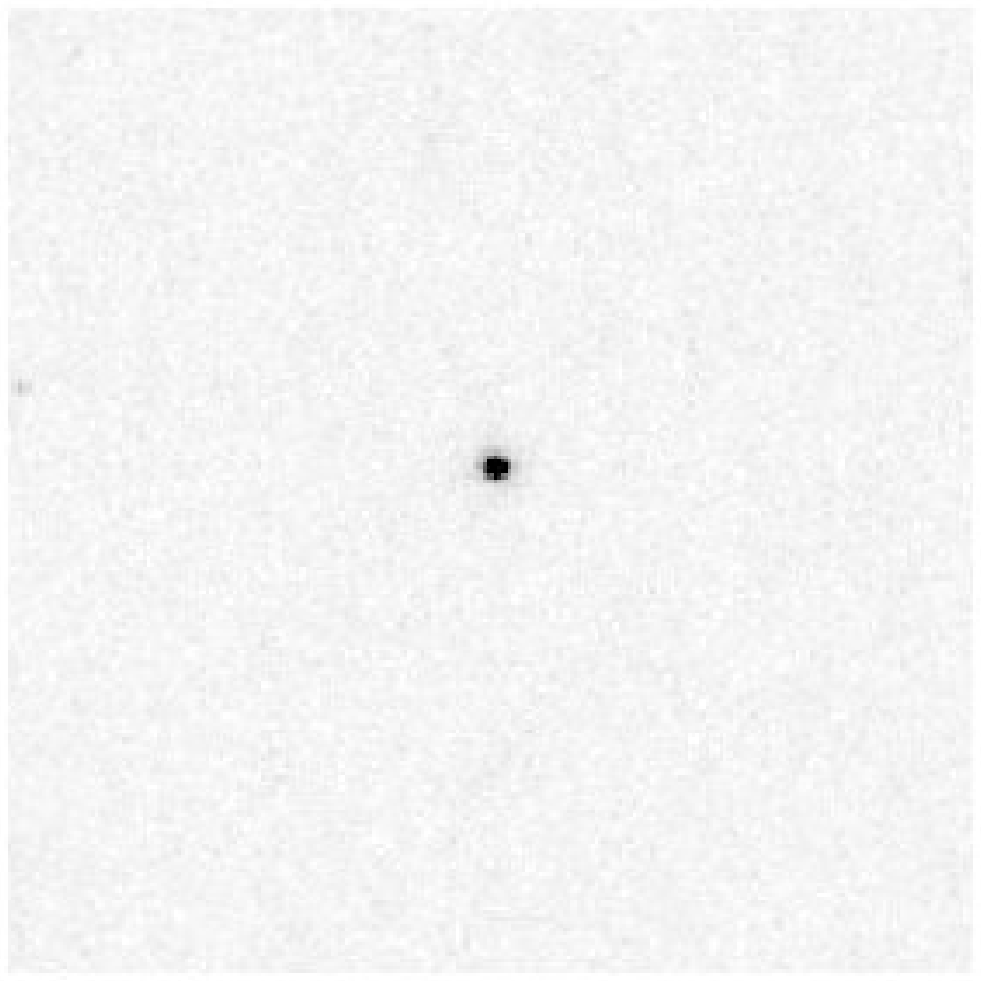}
\plotone{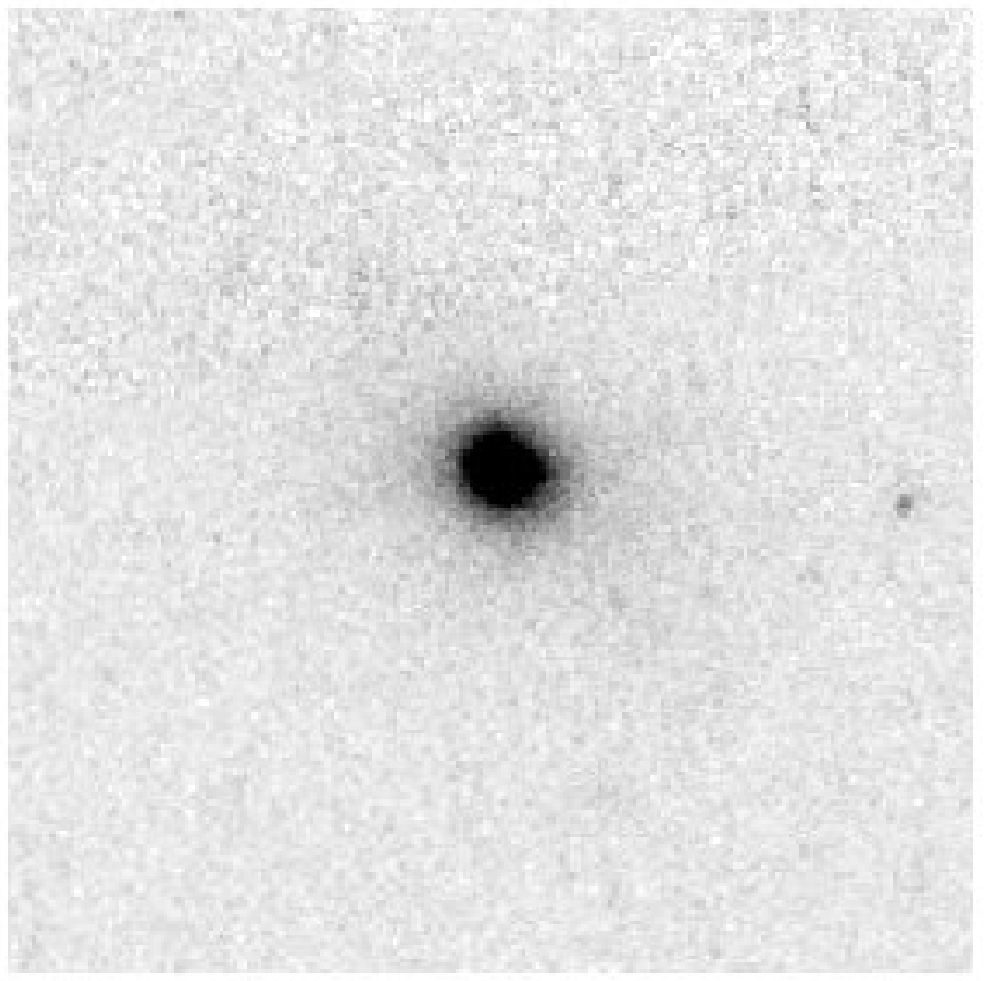}
\plotone{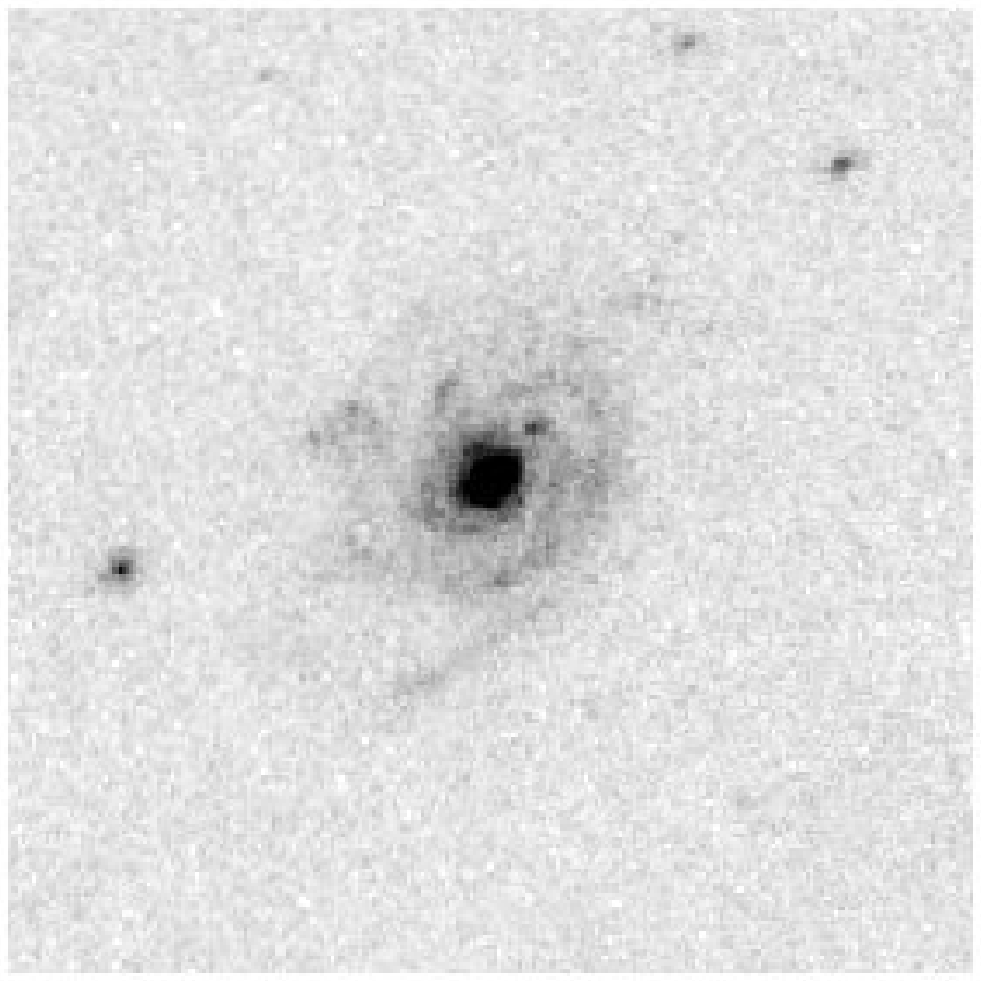}
\plotone{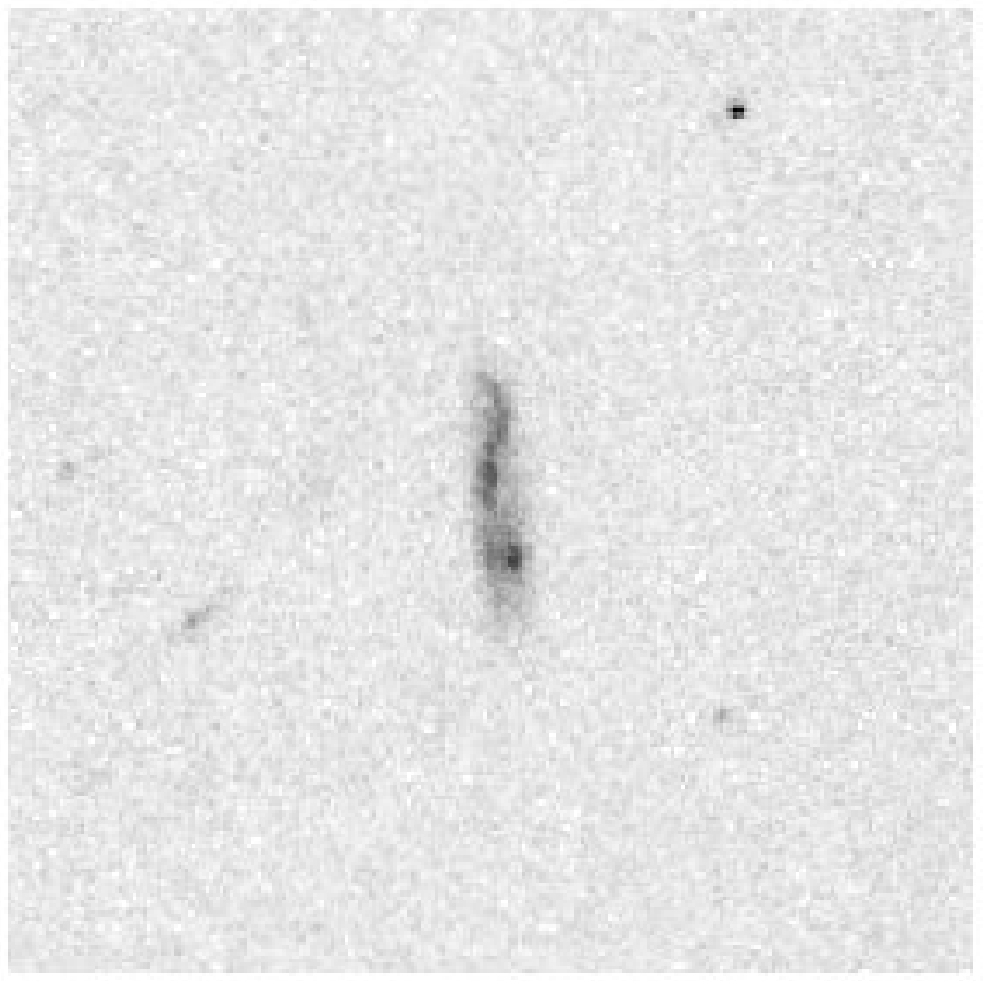}
\plotone{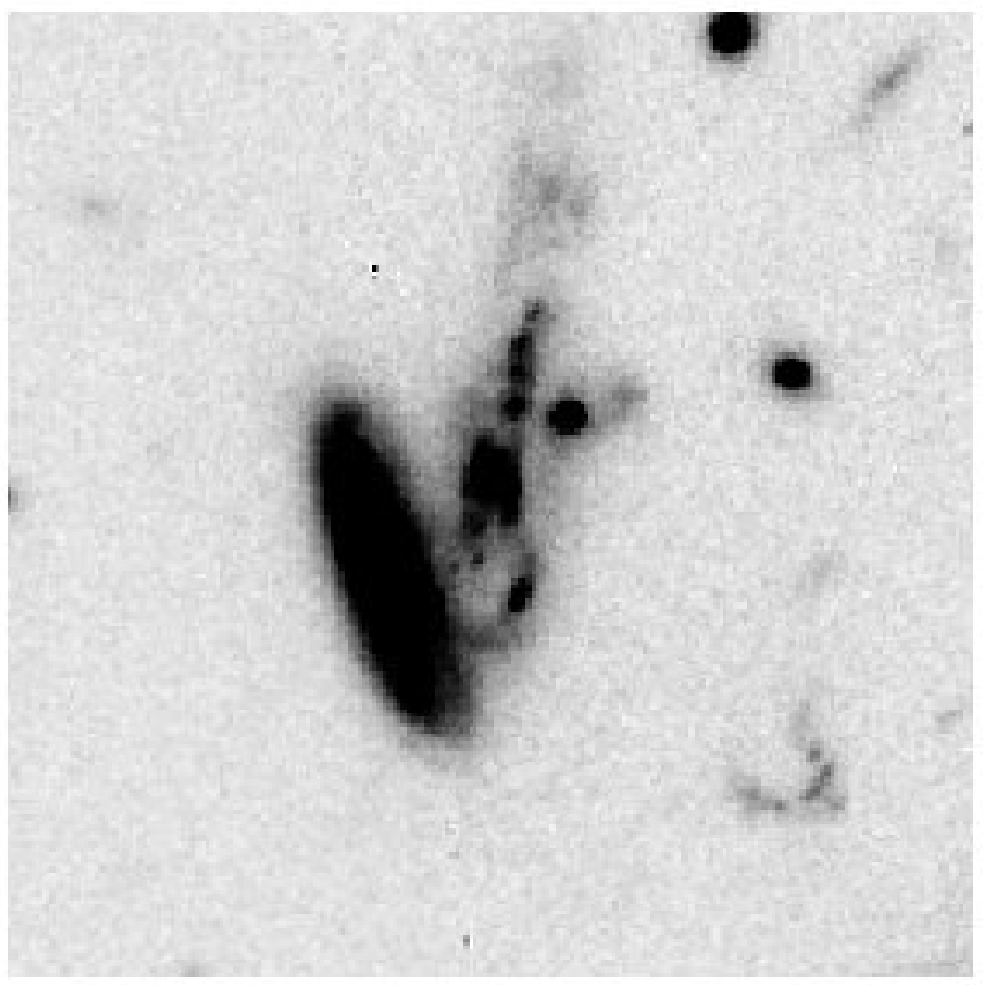}
\caption[Morphology]{Examples from {\it HST} ACS F814W of each type in
  our morphological classification; Compact, Elliptical, Spiral,
  Irregular/Disk, and Irregular/Merger.  All thumbnails are $10
  \arcsec$ on a side.  The Irregular/Merger example comes from the
  center of cluster 2 and is our only ULIRG.}
\label{fig:morph}
\epsscale{1}
\end{figure}

Since there are only 90 galaxies with MIPS detections at the cluster
redshift, we classify their morphologies manually by eye. For this, we
use the data with the best resolution which is the {\it HST} ACS F814W
data, corresponding to rest-frame B-band, with 0.5\arcsec/pixel
resolution. Training for this was done with examples from the online
SDSS GalaxyZoo\footnote{http://www.galaxyzoo.org/} which has color
images at a range of redshifts. We choose a very simple classification
scheme meant to divide those galaxies with visible signs of
interactions from those without. To this end we choose five categories
which fit all galaxies with the exception of seven galaxies because
they were either not imaged with ACS or are too near a bright star or
it's diffraction spikes to clearly classify. The five categories are
Compact, Elliptical, Spiral, Irregular/disk, and Irregular/merger. We
stick to very basic definitions to avoid ambiguous classifications.
Things fall into the compact, elliptical, or spiral classes if they
have classical forms of those shapes. Although compact classified
objects have the shape of a PSF, they have been confirmed to be
non-stellar based on their SED fits. Spirals include anything with a
disk that doesn't look disturbed or asymmetric in any way. Irregular
galaxies are anything that does not fit one of the classical
descriptions. Because the Irregular galaxies make up such a large
fraction of the sample, we have sub-divided that classification into
those systems that clearly have multiple nuclei or obvious tidal tails
(Irregular/merger) and all other irregular galaxies, mainly disturbed
disks (Irregular/disk). This differentiation of the irregular galaxies
may indicate something about the timescales of interaction histories
with the Irregular/merger classification going to those objects at
earlier stages of interaction, and Irregular/disk going to those
objects at later stages. Figure \ref{fig:morph} shows examples from
our sample of our morphological classification.

Table \ref{tab:morphtab} shows the morphological distribution for the
entire sample as well as subsamples based on color and infrared
luminosity (\S \ref{color} \& \ref{sfr}).  Of the entire sample of
member galaxies with MIPS detections, the majority of them are either
spirals or irregulars (81\%), unsurprisingly.  Specifically, 25\% of
the sources show obvious signs of interactions or mergers.  There are
potentially more interacting galaxies whose tidal features are too low
in surface brightness for us to detect but this cannot account for all
of them.  In the cases of the galaxies which show signs of interaction
we do not need to invoke a cluster environment driven process to
trigger star formation, we can assume here that the merger has
triggered star formation.  The remainder (75\%) of the sources which
do not show signs of interaction must have had their SF triggered by
some physical process that can occur within the cluster environment
such as ram pressure stripping or harassment \citep{gunn1972,
  moore1996}.  In the next section we discuss the location of this
SF to determine if it is on the cluster outskirts and therefore is
potentially residual SF after suppression upon entering the cluster
environment, or if it is truly being triggered by some cluster process
ongoing inside the cluster and suppression is not complete at the
cluster edges.

The majority of compact sources are part of the AGN sample based on
SED fitting.  Other compact determined galaxies are likely ellipticals
where the lower surface brightness outer parts of the bulge are not
visible at $z\sim1$.

The ellipticals are an interesting population in which to find star
formation.  From our original sample selection of 443 member galaxies,
less than two percent are ellipticals with MIPS detections.  About
half of the elliptical galaxies have red colors both before and after
extinction correction implying that there are a few legitimate red
ellipticals with star formation signatures.  Some of these are
possibly mis-classifications because of projections or surface
brightness dimming of a disk component or AGN mis-classifications.
Most of these ellipticals are sub-LIRGs so they do not have the higher
SFRs in the sample.  It is possible that we are seeing residual star
formation after a merger, but it is hard to imagine that the
morphological change would precede the end of the triggered star
formation.  The last possibility is that we see signs of dusty star
formation in elliptical galaxies that goes against traditional
findings that elliptical galaxies have no star formation, at least not
at the SFRs to which we are sensitive ($> 3\msun$).  Optically red,
morphologically elliptical galaxies with excess $24\micron$ emission
have also been found in SWIRE, GOODS, and the Bootes fields
\citep{rodighiero2007, davoodi2006, brand2009}.  While some of these
show AGN signatures, some are attributed to star formation.

When we split the sample based on infrared luminosity we see that the
spirals and irregulars make up the majority of the LIRGs (90\%) but a
lesser percentage of the sub-LIRGS (73\%) due to the higher fraction
of compact and elliptical sources.  Also interesting is that the
irregular population is split evenly between LIRGs and sub-LIRGs, and
60\% of the spirals are LIRGs.  In summary, LIRGs in clusters are most
likely to be blue spirals or irregulars.  Dividing by morphology,
spirals are more likely to be LIRGs, irregulars are equally likely to
be LIRGs or sub-LIRGs, and ellipticals are most likely sub-LIRGs.

Our findings of the ratios of morphological types in clusters is
similar to other published cluster values at high redshifts. In their
survey of a redshift 0.83 cluster, \citet{bai2007} find that of their
IR-detected galaxies, 20\%, 63\%, and 16\% of them are early-type,
late-type, and irregular galaxies, respectively, and 32\% show signs
of mergers/interactions. Also for a redshift 0.83 cluster,
\citet{marcillac2007} find 75\% spirals (including S0s, since we would
have given those a spiral designation) and 25\% irregulars. Again
these are only rough comparisons with the caveat that all of these
studies have only small samples which vary in cluster mass, density,
and dynamical state, etc., all things which might have an effect on
the morphologies and infrared luminosities of member galaxies.

\subsection{Distribution of Star Forming Galaxies}
\label{sfrdist}

\begin{figure}
\epsscale{0.4}
\plotone{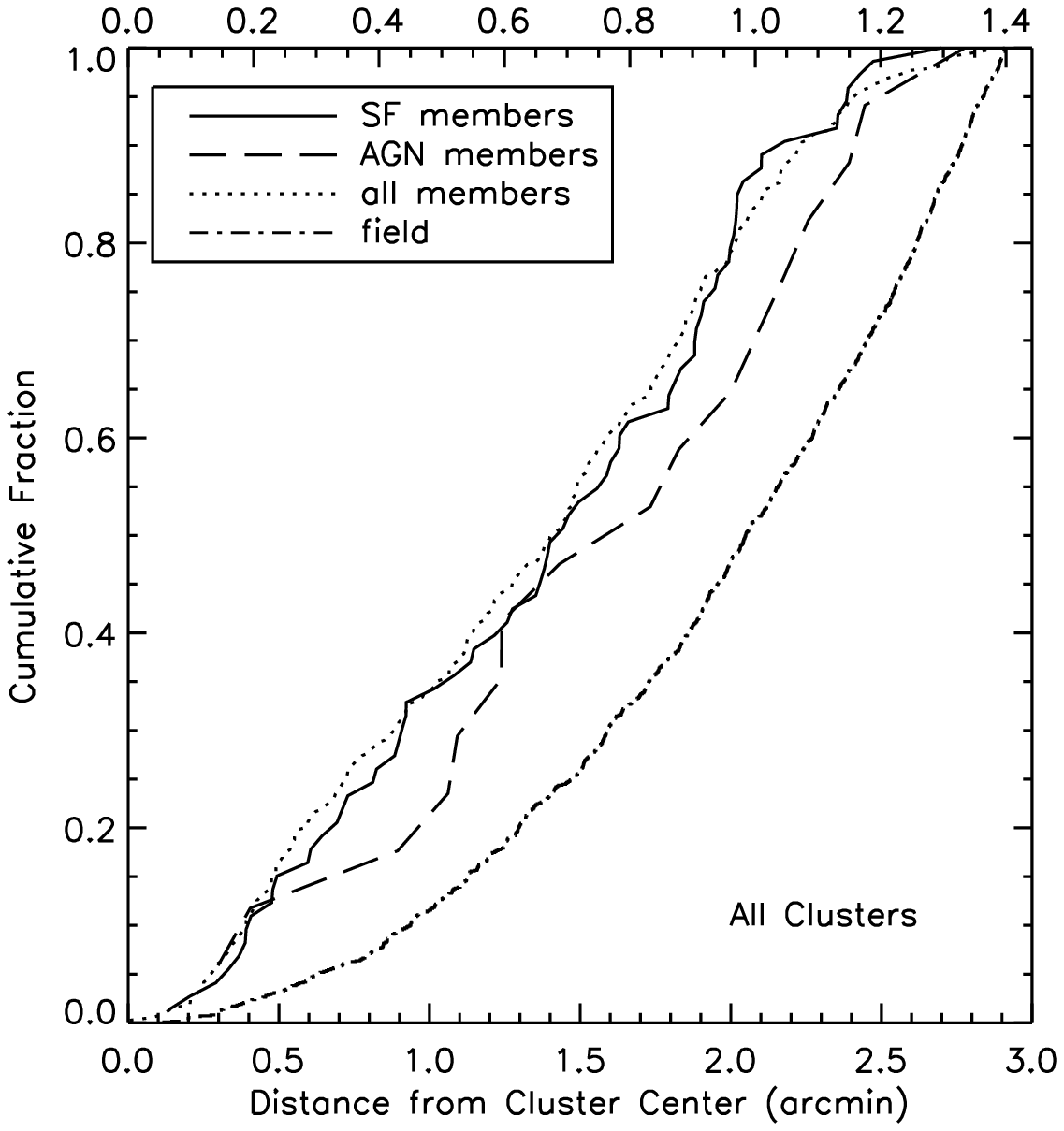}
\plotone{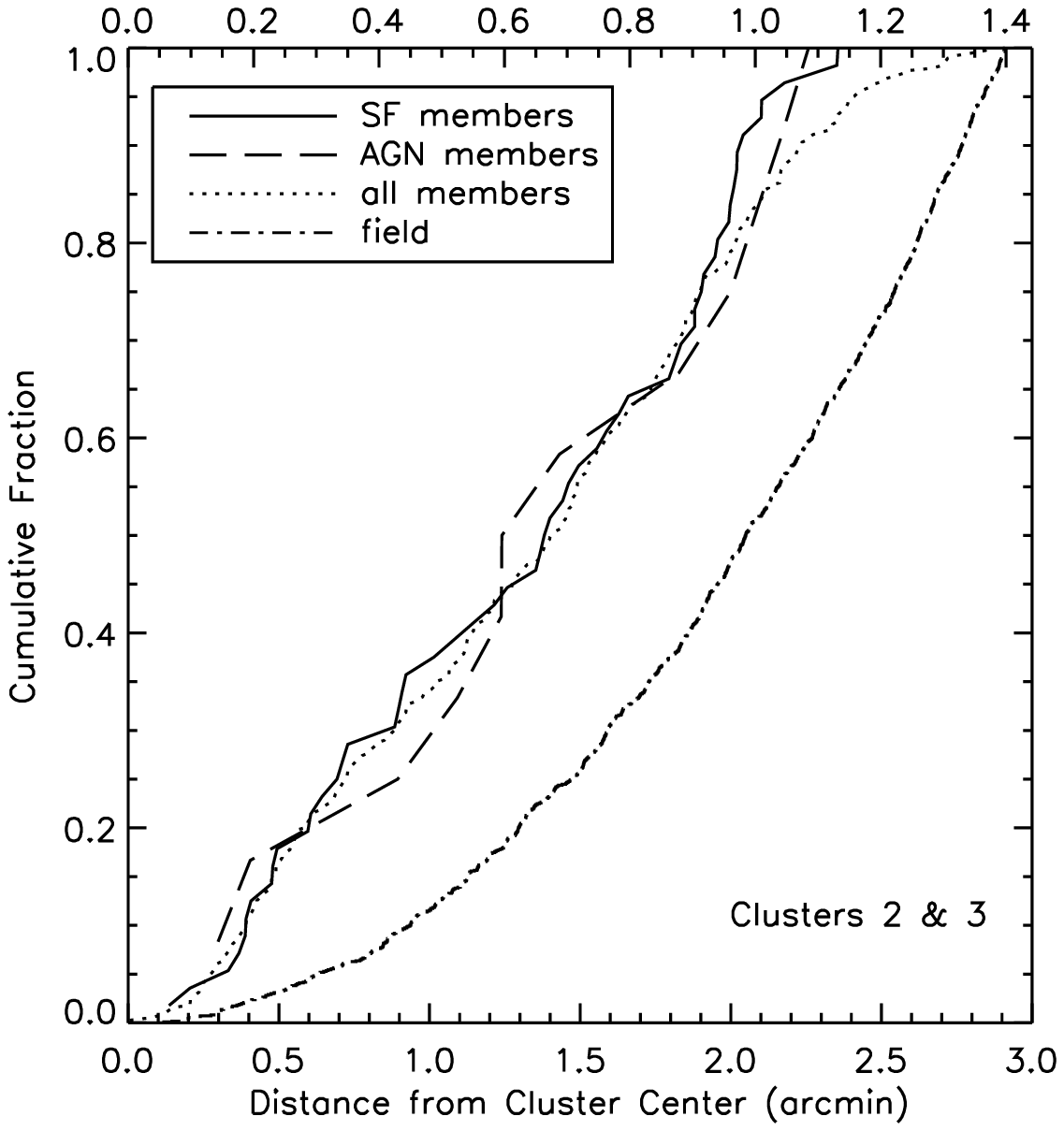}
\plotone{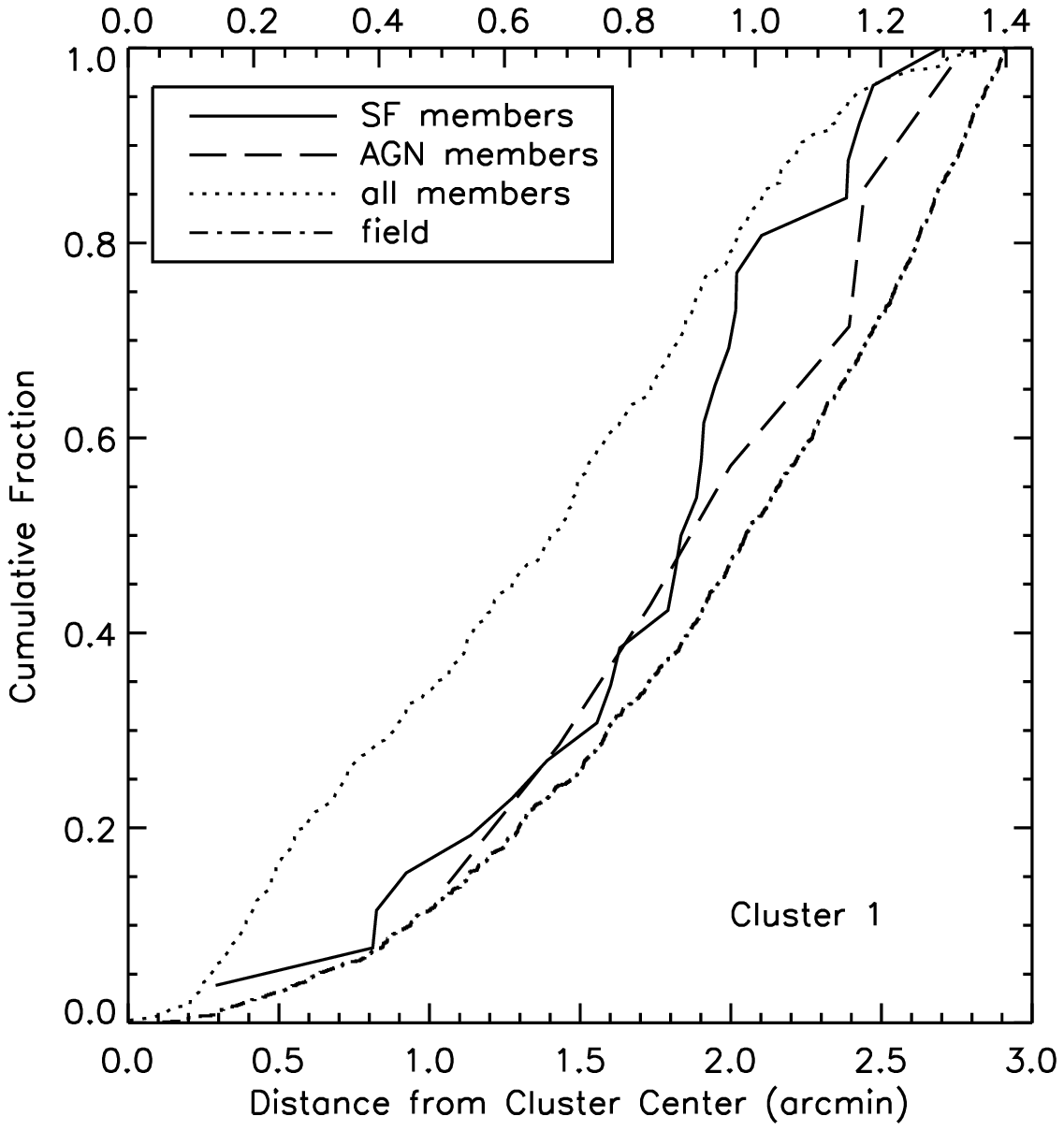}
\caption[cumdistspatial]{Cumulative distribution functions with
  distance from cluster center reported in arcminutes on the bottom
  axis and Mpc on the top axis. The solid, dashed, dotted, and
  dot-dashed lines represent the cluster star forming members, the
  cluster AGN members, all cluster members without MIPS flux or AGN
  SED shapes, and a field sample at redshift one respectively. {\it
    Top Left:} All three clusters combined; {\it Top Right:} Clusters
  2 and 3 only; {\it Bottom:} Cluster 1 only.}
\label{fig:cumdistspatial}
\epsscale{1}
\end{figure}

We examine the location of the MIPS sources in the clusters with the
goal of measuring if they are more or less concentrated than the
non-MIPS sources which would imply that they preferentially live in
the centers or outskirts of the clusters.  We make this comparison
using cumulative distributions and a KS test which is the most
straightforward way to determine if two continuous, unbinned,
distributions are drawn from the same parent distribution.  This is
the best statistical test to make this measurement given a relatively
small sample of galaxies especially when we split the sample by galaxy
property to examine the trends below.  KS tests are relevant on
samples sizes larger than $\sim 5$ \citep{press2007}.  Figure
\ref{fig:cumdistspatial} shows unbinned cumulative distributions as a function
of distance from the cluster center.  All three clusters are combined
here on the top left panel and distance from center is taken to be
distance to the nearest cluster center.  In the top left panel we show
the distribution of the MIPS-detected star forming galaxies (solid)
and AGN(dashed) compared to both all cluster members (dotted) and the
field(dot-dashed).  We check that increasing the sample to include
objects with 'only' five flux detections does not change the shape of
the cumulative distribution (see \S\ref{sample} for a discussion of
the number of detections required for an object to be included in the
sample).

The first thing to notice is that all of the cluster samples (those
with and without MIPS detections) show evidence of being significantly
more centrally concentrated than a comparison field sample as measured
in circles of the same area in the field.  A KS test on the composite
sample shows less than 1E-7\% chance that they are drawn from the same
parent population.  This is both nice confirmation of our photometric
redshifts and proof that star formation occurs in cluster environments.
In a similar experiment we determine the space density of MIPS sources
in the composite tri-cluster area compared to similar area in the
field.  In field regions of the same area as the cluster, we measure
the mean space density to be $43\pm15$ sources whereas we detect 90
sources in the cluster area which is a greater than $3\sigma$
overdensity. The cluster environment has clearly enhanced the number
of mid-IR sources among its member galaxies.  This is usually, but not
always the case in the literature.  \citet{geach2006} find only a very
minimal overdensity in MS0451-03 at z=0.55. \citet{marcillac2007,
  bai2007, gallazzi2008} all find significant overdensities when
compared to the field.

Secondly, the top left panel of figure \ref{fig:cumdistspatial} shows that
the star forming MIPS members and the non-MIPS-detected members are
consistent with having the same spatial distribution. A KS test shows
they have a 97\% chance of being drawn from the same population. 
We investigate this trend further by dividing our cluster sample.  The
top right and bottom panels of figure \ref{fig:cumdistspatial} show the
cumulative distributions for the separated clusters.  Interestingly,
cluster 1 on it's own has a significantly different spatial
distribution which has only a 1\% probability of being drawn from the
same population as the rest of the member galaxies.  In this cluster
we see a less concentrated distribution of star forming galaxies until
roughly 1 virial radius (0.7 Mpc) at which point the distribution steeply rises,
indicating a possible excess of star forming galaxies just beyond that radius.

The other two clusters show no such trend.  One possible explanation
is that there is some critical cluster property different between
these two sets driving the difference in spatial distributions.  One
could imagine that cluster property to be mass or evolutionary state.
Cluster one is both more massive than the other clusters and is more
relaxed in the sense that it appears to have already formed a cD
galaxy whereas the other clusters are in the process of forming their
cD's (see figure \ref{fig:morph} for an image of the central galaxy in
cluster 2).  A larger sample is required to examine these differences.
A second possibility is that these two clusters represent a complex
structure in our 2D image.  They are relatively close to each other
(overlapping virial radii at the same photometric redshifts) that it is
possible these two clusters actually reside in the same potential
well, or that one is falling in towards the center of the other, which
would make our choice of centers meaningless.  Because of their nearness, we
could imagine that projection effects could dilute any potential
signal of a less concentrated distribution.

We have found that different clusters potentially exhibit different
spatial distributions in their star forming galaxies, which is also
found in \citet{geach2006}.
\citet{coia2005,bai2007,marcillac2007,gallazzi2008,fadda2008} and
\citet{ koyama2008} report the detection of an intermediate density at which
cluster star forming galaxies congregate, but this is also not found
in the work of \citet{biviano2004}.  The comparison of literature
samples is not straightforward because of the differing cluster
properties (mass, virialization, and structure) and differing sampling
methods including flux detection levels and accounting for AGN
contamination.  Also in some cases the evidence for star forming
galaxies to preferentially lie at intermediate densities is not
statistically strong ($< 3\sigma$).  For these listed reasons, and
that different authors use different measures of local environment, it
is not practical to compare literature samples.

We further discuss the cumulative distributions of the sample with a
focus on infrared luminosity, morphology, and color.  We continue
to discuss the sample as the combination of all three clusters which
does not effect the remainder of the results.

\subsubsection{Distribution by L$_{IR}$}
\label{lirdistsect}
We divide the sample of star forming galaxies based on infrared
luminosity in the top right panel of Figure \ref{fig:cumdist}. Those
with LIRG luminosities or above are shown with the solid line, and
those with sub-LIRG luminosities are shown with the dotted line. The
LIRGs do appear to be more centrally concentrated than the sub-LIRGS,
however a KS test is inconclusive giving a 53\% probability that they
are drawn from the same population. This inconclusiveness means that
we cannot rule out the possibility that sub-LIRGs have a different,
less concentrated, distribution than LIRGs. This leaves open the
possibility that at lower redshift where surveys are likely to be
deeper than high redshift surveys, the lower luminosity sub-LIRGS
might dominate the population thereby giving the appearance of being
overall less concentrated than the other member galaxies. This could
be a reason why lower redshift surveys find less concentrated spatial
distributions, but does not explain the preferred density peaks
reported in those studies.

\subsubsection{Distribution by Morphologies}
\label{morphdistsect}

We now investigate the location of the star forming galaxies by
splitting the sample on morphology. If the spirals are less centrally
concentrated it could suggest that the cluster environment is able to
burst and than suppress star formation in normal non-interacting
galaxies.  The remaining star formation activity that we see closer to
the center is then the result of galaxy interactions.  The top right
panel of Figure \ref{fig:cumdist} shows the cumulative distribution of
the spiral sample (solid line) and the likely merger sample (dotted
line), based on the morphologies as determined by eye in \S
\ref{morph}.  There is again tantalizing but inconclusive evidence
that the spirals are less centrally concentrated.  A KS test on these
two samples gives a 65\% chance that the two populations draw from the
same parent distribution which prevents us from concluding either that
there are or not more centrally concentrated.  We also plot the
distribution of the elliptical star forming galaxies but small sample
size (7 galaxies) prevents us from making conclusions.  Projection
effects also complicate this analysis since we do not know the 3D
location of the member galaxies.

The inconclusive tests for both morphology and L$_{IR}$ radial
distributions are probably telling us that there is another variable
which is confusing these tests.  A larger sample size of clusters
split by cluster properties is desirable to further test our
hypotheses in these cases.
\subsubsection{Distribution by Color}
\label{colordistsect}

\begin{figure}
\epsscale{0.4}
\plotone{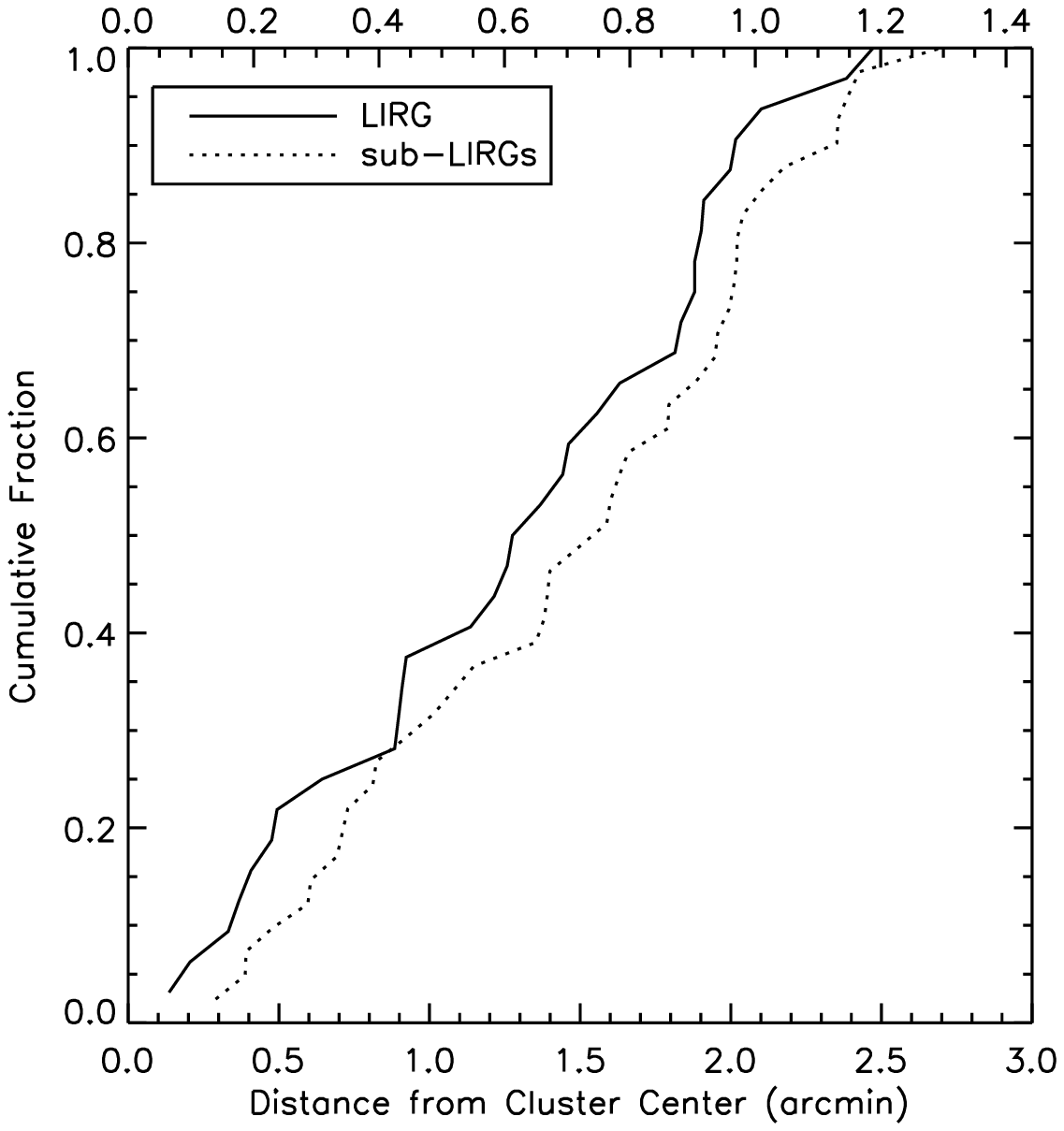}
\plotone{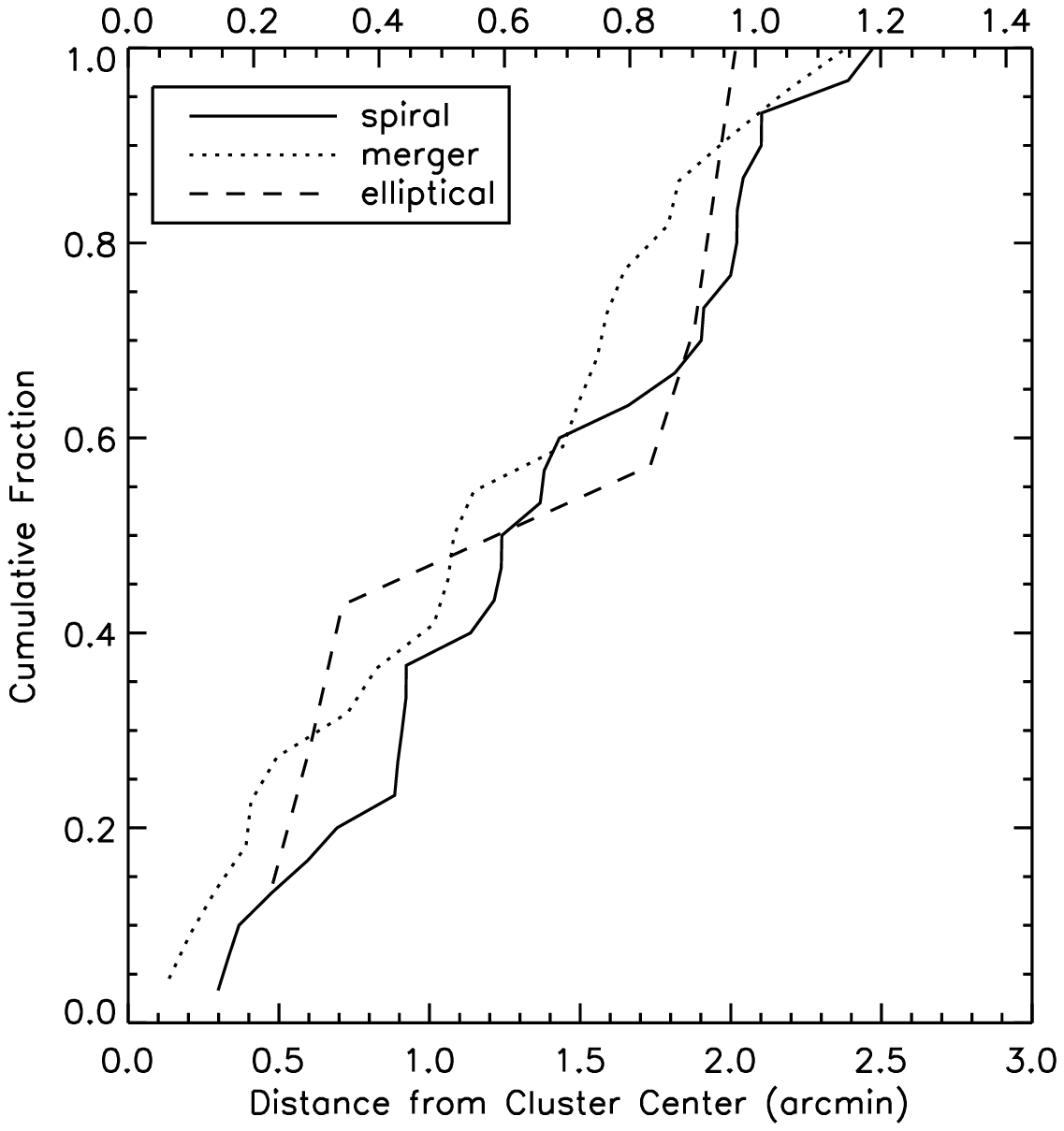}
\plotone{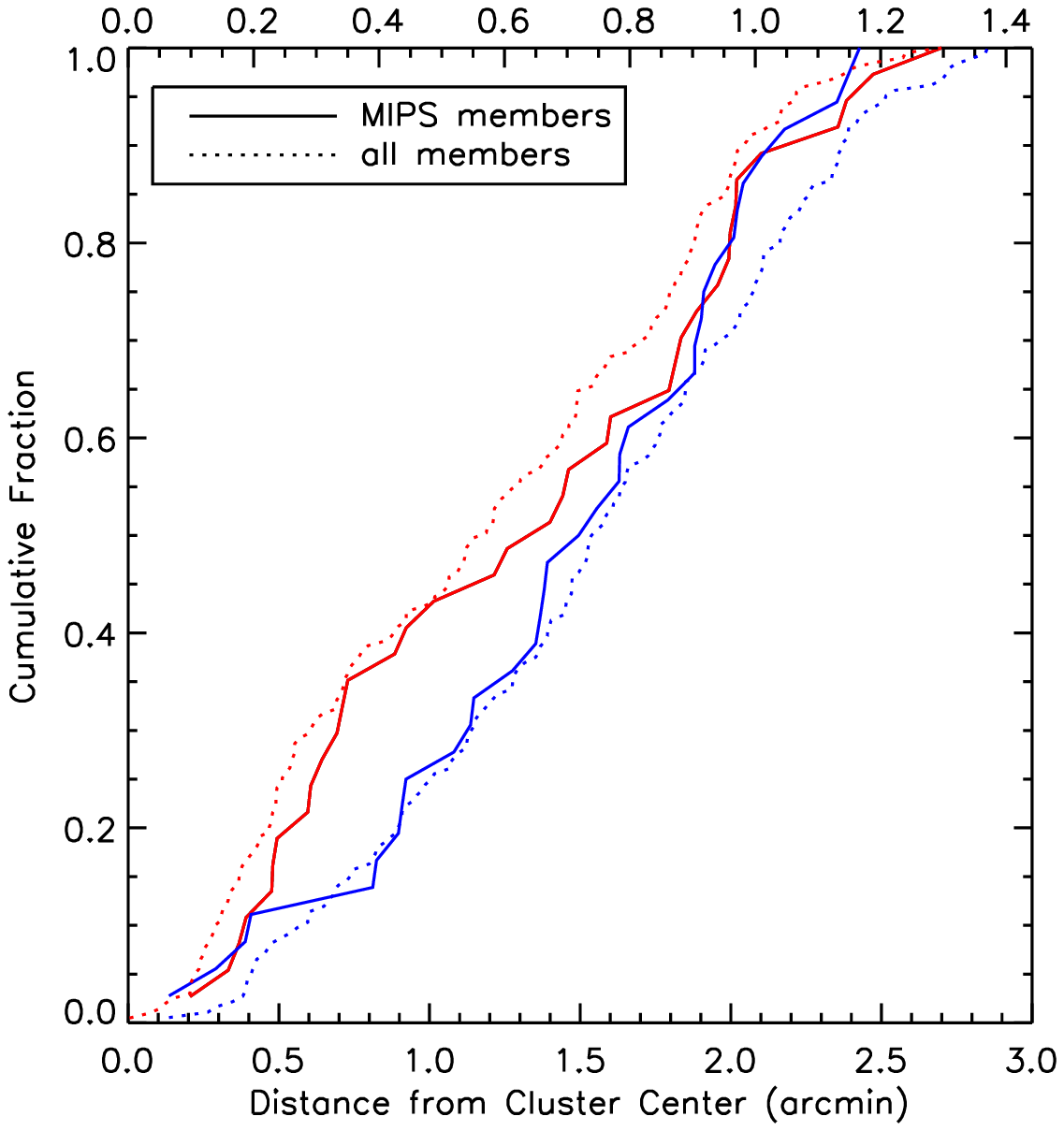}
\plotone{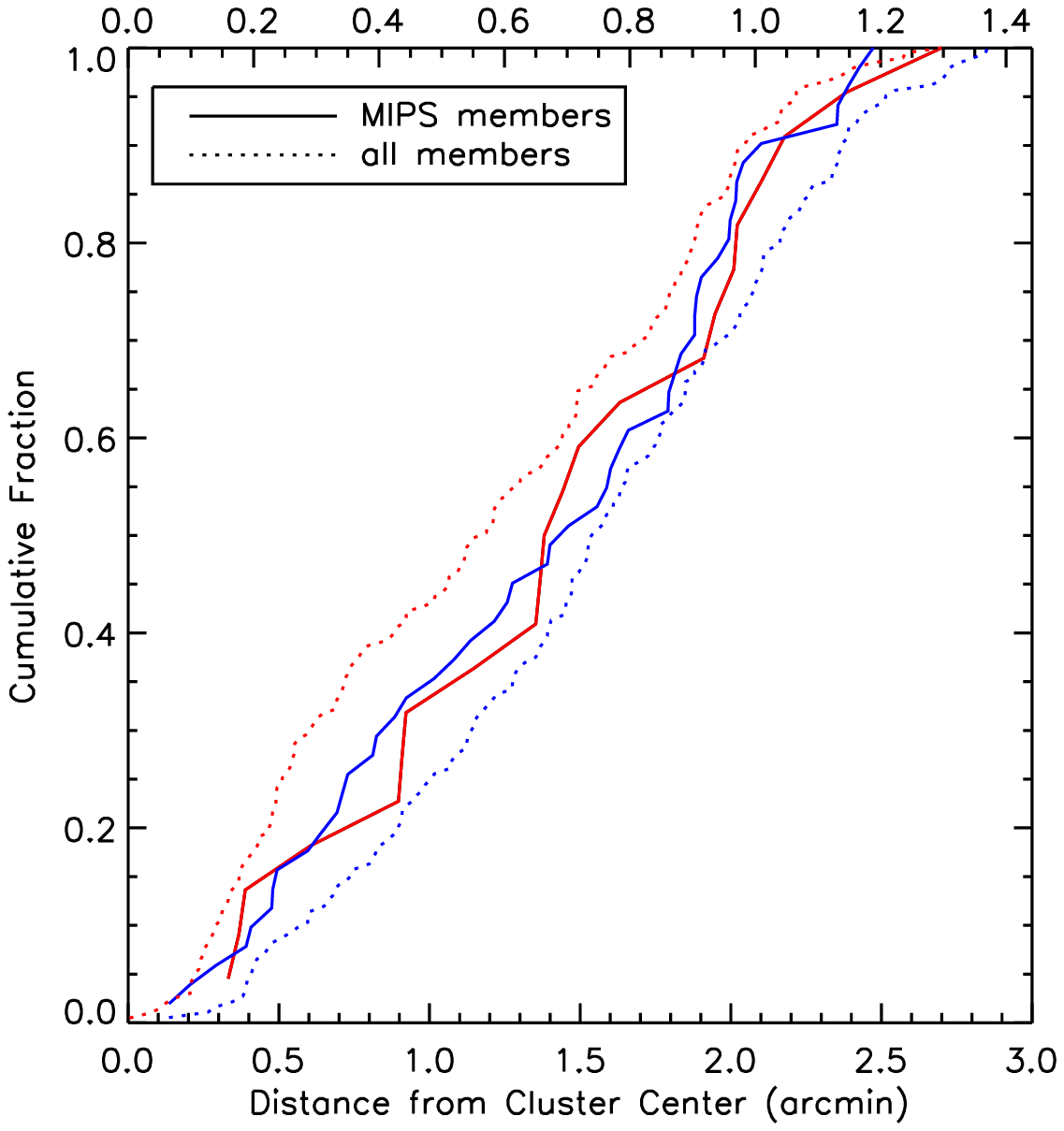}
\caption[cumdist]{Cumulative distribution functions with distance from
  cluster center reported in arcminutes on the bottom axis and Mpc on
  the top axis.  {\it Top left} Distribution split by infrared
  luminosity.  LIRGS (including the lone ULIRG) are shown with the
  solid line, while sub-LIRGS are shown with the dotted line.  {\it
    Top right} Distribution split by morphology into spirals (solid) ,
  ellipticals (dashed), and irregular/mergers (dotted). {\it Bottom
    left} Distribution split by color, uncorrected for reddening.  The
  solid and dotted lines represent the cluster star forming members
  and all cluster members without MIPS flux or AGN SED shapes.  The
  more concentrated set of solid and dotted (red) lines represent the
  red galaxies while the less concentrated (blue) set of lines show
  the blue galaxies. {\it Bottom right} Distribution split by
  reddening corrected color; same line definitions as the middle right
  plot. }
\label{fig:cumdist}
\epsscale{1}
\end{figure}

To understand more about spatial distribution of the MIPS sources we
further divide the sample of star forming galaxies by color into a red
and blue sample based on their uncorrected magnitudes.  The dividing
line is taken to be the blue edge of the red cluster sequence as
described in paper one and shown in Figure \ref{fig:colordist}.  The
same division is made for the non-MIPS-detected cluster members and
the results are shown in the bottom left panel of Figure
\ref{fig:cumdist}.  The left, more centrally concentrated fork of the
distribution shows the red galaxies while the right fork shows the
distribution of the blue galaxies with solid lines for the MIPS
members and doted lines for the non-MIPS member galaxies.  Again we
see that the MIPS members and non-MIPS members show similar
distributions (KS =76\& for red and 70\% for blue) while we see a
clear difference between red and blue galaxies (KS $< 0.01\%$) with
blue avoiding the central dense regions of the clusters.  This is a
classic finding that blue galaxies generally don't inhabit dense
environments \citep{butcher1984, pimbblet2003}.

However, if we look at the distribution of the reddening corrected
colors (\S\ref{color}) we find a different story; bottom right panel
of Figure \ref{fig:cumdist}. Here we see that the blue non-MIPS
-detected members still show the same trend of the blue galaxies
avoiding the centers. In contrast to the non-reddening corrected color
distributions, all the MIPS-detected galaxies now have the same
distribution regardless of color (KS = 94\%).  In other words, the formerly red
galaxies are co-spatial with the blue galaxies. This is just showing
us again that many of the observed red galaxies are actually dusty
blue galaxies and are not red because they are old.

\subsection{Distribution of AGN}
\label{agndist}

We examine the distribution of cluster AGN compared to the MIPS
-detected star forming galaxies. Figure \ref{fig:cumdistspatial}
includes the cumulative distribution of MIPS -detected AGN cluster
members as the dashed lines. A KS test between the AGN and star
forming members in the combined distribution is inconclusive, showing
a 50\% probability of drawing from the same population. A KS test on
clusters two and three shows a 90\% probability of deriving from the
same population. It is interesting that the AGN and star forming
galaxies appear to have similar distributions in two of the clusters.
The similar distribution could imply that the same physical mechanism
triggers AGN and star formation. In a similar case to the SF galaxies,
because cluster one shows a different AGN distribution from the other
clusters, we are unable to ferret out the underlying causes of the
distributions. The literature is similarly inconclusive. A radio
sample from \citet{lin2007} shows that AGN are more concentrated than
cluster galaxies with the radio-brightest being the most concentrated.
A sample of eight low-intermediate redshift clusters with
X-ray-detected AGN reveal the same trend \citep{martini2007}. However,
in a supercluster at z=0.9, \citet{kocevski2008} find that X-ray AGN
are more likely located in the intermediate regions, avoiding the
densest cluster centers. The differences in samples between Radio,
X-ray, and Mid-IR selections and differences between depths and
cluster characteristics may be the source of these differences. A
larger sample is necessary to make progress on this topic.



\section{Conclusion}
\label{conclusion}

We have used a multi-wavelength dataset based on extremely deep {\it
  Spitzer} IRAC data to examine the nature of mid-IR sources in a
large scale structure of three clusters at redshift one.  There are 90
members of the clusters with MIPS detections within two virial radii
of the cluster centers, of which 17 appear to have SEDs dominated by
AGN and the rest dominated by star formation.  With the samples of AGN
and star forming sources we examine the total infrared luminosities,
star formation rates of individual galaxies and of the structure as a
whole, colors, morphologies, and distributions whereby we come to the
conclusions listed below.

\begin{itemize}

\item We look for evolution in the AGN fraction with redshift.  In a
  comparison with X-ray surveys we find a continued increase in the
  AGN fraction out to redshift one with trepidation over the accuracy
  of the conversion between $L_{5.8} $ and $L_{X}$.  In addition the
  magnitude of the increase in AGN fraction is higher in clusters than
  in the field.  If an effect of AGN activity is to suppress star
  formation through a feedback mechanism, then the measured large
  number of AGN at higher redshifts indicates that there will be many
  galaxies for which AGN feedback may be a significant player in
  turning off star formation in lower redshift clusters.  Secondly
  because of the more rapid decrease in AGN fraction in clusters
  compared to the field, we conclude that cluster environment has an
  effect on the decline of the AGN population.

\item For the sample of star forming members, we use the 24\micron\
  flux (rest frame 12\micron)to estimate total infrared luminosity.
  The distribution of infrared luminosities shows that about half of
  the sample have infrared luminosities consistent with being LIRGs
  while the other half are sub-LIRGs.  That distribution is consistent
  with the field at redshift one, as measured from other regions in
  our data, implying that the cluster environment does not have an
  effect on the infrared luminosities of the galaxies within it.

\item Total infrared luminosity is converted to star formation rate.
  As a whole, the summed, mass-normalized cluster star formation rate
  is higher at $z=1$ than in counterparts at lower redshift.  The measured
  decrease of SFRD from z=1 to 0 is larger than the decrease measured
  in the field implying that suppression of star formation is
  accelerated in the cluster environment.

\item Based on SED fitted extinction values at rest frame B-band, we
  find that MIPS sources in clusters are mainly highly extincted,
  dusty, intrinsically blue galaxies.  A few are intrinsically old red
  galaxies.

\item Morphologies of the MIPS-detected sources are determined by eye
  from the {\it HST} rest-frame B-band images.  The majority of
  sources (81\%) are spirals or irregulars.  There are a few
  elliptical galaxies (8\%) , the majority of which have sub-LIRG
  luminosities.  Potentially some of these are mis-classifications,
  but some are real detections of dusty star formation of greater than
  three solar masses per year in an elliptical galaxy.  The LIRGs in
  clusters are most likely to be blue spirals or irregulars.  A large
  fraction (at least 25\%) show obvious signs of interactions.  This
  implies that some cluster galaxies have SF triggered by the cluster
  environment and not solely by merger processes which are not cluster
  environment dependent.

\item Cluster MIPS sources are significantly more concentrated than a
  field sample at redshift one showing that they are indeed members of
  the cluster.  Cluster characteristics appear to influence the spatial
  distribution of the star forming member galaxies.  One of our
  clusters shows the MIPS sources with a less concentrated
  distribution than the other members.  However, the other two
  clusters have MIPS sources with the same distribution as the member
  galaxies implying that complete suppression has not occured due to the
  cluster environment.  There is inconclusive evidence for LIRGs and
  irregular galaxies separately to be more centrally concentrated than
  sub-LIRGs and spirals respectively.  When using uncorrected
  magnitudes, galaxies blue-ward of the red sequence are significantly
  less concentrated than red galaxies.  However when using reddening
  corrected galaxy colors, we find all MIPS-detected cluster members
  to have the same distribution confirming that the MIPS sources
  really are dusty, star forming, blue galaxies and not a separate
  population.

\end{itemize}
Cluster environment does seem to have an effect on the evolution of
AGN fraction and SFR from redshift one to the present, but amongst the
IR active galaxy sample, environment does not affect the infrared
luminosities. This may be saying that whatever triggers the star
formation in clusters has the same effect on the galaxies in the
clusters as whatever triggers star formation in the field, eg. star
formation looks the same regardless of environment. Or, in other
words, the effect of star formation on a galaxies infrared luminosity
is independent of triggering mechanism. But, the cluster environment
does encourage SFR and AGN fraction to decline more rapidly with time
over the field implying that the cluster environment does have an
effect on a galaxies activeness, either SF or AGN. While some of our
galaxies show signs of interaction as the likely triggering mechanism,
it seems likely that other cluster environment driven effects are also
able to trigger SF within the cluster. This is based both on
morphological indicators of SF and the distributions of SF galaxies.
In two clusters we see no evidence for a suppression of star formation
in the inner regions of the clusters as we would expect if there were
a density cutoff for triggering SF. As always, a larger sample of
clusters with deeper mid-IR measurements is desirable.


\acknowledgments

This research has made use of data from the Two Micron All Sky Survey,
which is a joint project of the University of Massachusetts and the
Infrared Processing and Analysis Center/California Institute of
Technology, funded by the National Aeronautics and Space
Administration and the National Science Foundation.  This work was
based on observations obtained with the Hale Telescope, Palomar
Observatory as part of a continuing collaboration between the
California Institute of Technology, NASA/JPL, and Cornell University,
the {\it Spitzer} Space Telescope, which is operated by the Jet Propulsion
Laboratory, California Institute of Technology under a contract with
NASA, the MMT Observatory, a joint facility of the Smithsonian
Institution and the University of Arizona, and the NASA/ESA Hubble
Space Telescope, obtained at the Space Telescope Science Institute,
which is operated by the Association of Universities for Research in
Astronomy, Inc., under NASA contract NAS 5-26555. These observations
are associated with program \#10521.  Support for program \#10521 was
provided by NASA through a grant from the Space Telescope Science
Institute, which is operated by the Association of Universities for
Research in Astronomy, Inc., under NASA contract NAS 5-26555.

{\it Facilities:} \facility{Hale (LFC, WIRC, COSMIC)}, \facility{MMT
  (Megacam)}, \facility{HST (ACS)}, \facility{Spitzer (IRAC, MIPS)},
\facility{Akari}, \facility{CXO (ACIS)}

\bibliography{ms.bbl}  
\clearpage

\begin{deluxetable}{ccccccc}
\tablewidth{0pc}
\tablecolumns{7}
\tablecaption{Cluster Characteristics \label{tab:clusterchar}}

\tablehead{
\colhead{Cluster} &         
\colhead{ra} &        
\colhead{dec} &      
\colhead{$N_{gals}$} & 
\colhead{z$_{peak}\tablenotemark{a}$} & 
\colhead{L$_{x}$ (0.5-2.0 Kev)} &    
\colhead{M$_{500}$}
\\
\colhead{ } &
\colhead{J2000 (deg)} &      
\colhead{J2000 (deg)} &  
\colhead{$r < 500 Kpc$} &
\colhead{ }            & 
\colhead{$1\times10^{43}$erg/s} & 
\colhead{$1\times10^{13}\msun$}
}

\startdata
         1 &  264.68160 &   69.04481 &        215 & $1.0\pm0.1$ & $3.6\pm0.6$ & $6.2\pm1.4$ \\

         2 &  264.89228 &   69.06851 &        255 & $1.0\pm0.1$ & $1.6\pm0.7$ & $3.6\pm1.4$ \\

         3 &  264.83102 &   69.09031 &        241 & $1.0\pm0.2$ & $\le1.6\pm0.7$ & $\le3.6\pm1.1$ \\

\enddata
\tablenotetext{a}{Redshift peak and one sigma uncertainty are measured from a Gaussian fit to the redshift distribution.}
\end{deluxetable}  

\begin{deluxetable}{cccc}
\tablewidth{0pc}
\tablecolumns{7}
\tablecaption{Morphologies of MIPS24 members \label{tab:morphtab}}

\tablehead{
\colhead{Galaxy Morphology} &         
\colhead{All} &        
\colhead{LIRGs} &      
\colhead{sub-LIRGs}  
}

\startdata
         Compact     &  {\bf 9 (11\%)}    &2 (5\%)       & 7 (16\%)\\

         Elliptical     &  {\bf 7 (8\%)}     & 2 (5\%)     & 5 (11\%)\\

         Spiral         &  {\bf 30 (35\%)}  & 18( 44\%) & 12 (27\%)\\

         Irr/Disk     &  {\bf 18 (21\%)}     &9 (22\%)   & 9 (21\%)\\

         Irr/Merger &  {\bf 21 (25\%)}     & 10 (24\%) & 11 (25\%)\\

 \tableline \\
         Total          &{\bf 85 (100\%) } & 41 (100\%) & 44 (100\%) \\
\enddata

\tablecomments{The first data column shows
  the morphology breakdown for all member galaxies with MIPS
  detections.    Columns 2 \& 3 divide all members into those with
  LIRG and sub-LIRG luminosities.  Percentages are of the galaxies
  only within the column shown.}
\end{deluxetable}  





\end{document}